\documentclass[aps,twocolumn,prl]{revtex4-1}
\usepackage{graphicx,amsmath, amsfonts,bm, color}

\usepackage{xcolor,hyperref}
\hypersetup{
   colorlinks,
   linkcolor={blue!50!black},
   citecolor={blue!50!black},
   urlcolor={blue!80!black}
}

\usepackage{enumitem}

\renewcommand{\=}{\!=\!}
\newcommand{\1}{^{\mbox{\tiny (1)}}}

\DeclareMathOperator{\sgn}{sgn}
\newcommand{\Vst}{\ensuremath{V_\text{stick}}}
\newcommand{\Gb}{\ensuremath{\overline{G}}}
\newcommand{\dbar}{{\,\mathchar'26\mkern-12mu d}}

\begin{document}

\title{Unstable slip pulses and earthquake nucleation as\\ a non-equilibrium first-order phase transition}
\author{Efim A.~Brener$^{1}$}
\thanks{E.~A.~Brener and M.~Aldam contributed equally.}
\author{Michael Aldam$^{2}$}
\thanks{E.~A.~Brener and M.~Aldam contributed equally.}
\author{Fabian Barras$^{3}$}
\author{Jean-Fran\c{c}ois Molinari$^{3}$}
\author{Eran Bouchbinder$^{2}$}

\affiliation{$^{1}$Peter Gr\"unberg Institut, Forschungszentrum J\"ulich, D-52425 J\"ulich, Germany\\
$^{2}$Chemical and Biological Physics Department, Weizmann Institute of Science, Rehovot 7610001, Israel\\
$^{3}$Civil Engineering Institute, Materials Science and Engineering Institute, Ecole Polytechnique F\'ed\'erale de Lausanne, Station 18, CH-1015 Lausanne, Switzerland}

\begin{abstract}
The onset of rapid slip along initially quiescent frictional interfaces, the process of ``earthquake nucleation'', and dissipative spatiotemporal slippage dynamics play important roles in a broad range of physical systems. Here we first show that interfaces described by generic friction laws feature stress-dependent steady-state slip pulse solutions, which are unstable in the quasi-1D approximation of thin elastic bodies. We propose that such unstable slip pulses of linear size $L^*$ and characteristic amplitude are ``critical nuclei'' for rapid slip in a non-equilibrium analogy to equilibrium first-order phase transitions, and quantitatively support this idea by dynamical calculations. We then perform 2D numerical calculations that indicate that the nucleation length $L^*$ exists also in 2D, and that the existence of a fracture mechanics Griffith-like length $L_G\!<\!L^*$ gives rise to a richer phase-diagram that features also sustained slip pulses.
\end{abstract}

\maketitle

{\em Introduction.---} The spatiotemporal dynamics of frictional interfaces (``faults''), formed when two deformable bodies come into contact, are central to a broad range of physical systems~\cite{Baumberger2006,Ben-Zion2008,Vanossi2013}. Two basic recurring themes, which still resist a complete theoretical understanding, are rapid slip nucleation and the rupture modes of faults. The former addresses the conditions under which slowly driven or strictly quiescent faults spontaneously develop rapid slip, the so-called ``earthquake nucleation'' problem~\cite{Ruina1983,Yamashita1991,Ben-Zion1997,Ben-Zion2001,Ampuero2002,Lapusta2003,Uenishi2003,Rubin2005,Ampuero2008,Kaneko2008,McLaskey2013,Latour2013,Viesca2016, Viesca2016b,Kaneko2016,Kaneko2017,aldam2017a,Gabriel2012}. The latter addresses to the ways in which such faults rupture once rapid slip nucleates, in particular the existence and properties of expanding crack-like rupture vs.~spatially-compact pulse modes~\cite{Gabriel2012,Perrin1995,Beeler1996,Cochard1996,Zheng1998,Nielsen2000,Bizzarri2003,Brener2005,Rubin2009,Ben-David2010a,Bar-Sinai2013,Svetlizky2014,Putelat2017}.

Earthquake nucleation has been extensively studied~\cite{Ruina1983,Yamashita1991,Ben-Zion1997,Ben-Zion2001,Ampuero2002,Lapusta2003,Uenishi2003,Rubin2005,Ampuero2008,Kaneko2008,Gabriel2012, McLaskey2013,Latour2013,Viesca2016,Viesca2016b,Kaneko2016,Kaneko2017,aldam2017a}. It has been shown that for a broad class of interfaces where the frictional resistance decreases with increasing slip velocity, i.e.~in the velocity-weakening regime, nucleation emerges from a frictional instability~\cite{Ruina1983,Yamashita1991,Ben-Zion1997,Ben-Zion2001,Ampuero2002,Lapusta2003,Uenishi2003,Rubin2005,Ampuero2008,Kaneko2008,McLaskey2013,Latour2013,Viesca2016, Viesca2016b,Kaneko2016,Kaneko2017,aldam2017a,Gabriel2012}. This nucleation scenario, controlled by a critical nucleation length $L_{\rm c}$~\cite{SM}, assumes that external driving forces bring the interface or part of it into the destabilizing velocity-weakening regime, which is valid only above some typically low slip velocity. Far less is known about nucleation from the quiescent, nearly locked state that is generically velocity-strengthening~\cite{Hardebeck2018}.

Once rapid slip commences, the spatiotemporal dynamics of frictional interfaces are largely determined by the mode of rupture propagation along them, e.g.~\cite{Ben-Zion2001,Ben-Zion2008}. While expanding crack-like rupture has been thought to be the dominant mode of rupture, it has been suggested that some earthquake data might be explained in terms of slip pulses~\cite{Heaton1990}. This suggestion has triggered various 2D analyses~\cite{Perrin1995,Beeler1996,Cochard1996,Zheng1998,Nielsen2000} that demonstrated the existence of slip pulses for a class of friction models that feature aging/healing in the absence of slip and sufficiently strong velocity-weakening behavior. Recently, the existence of steady-state slip pulses in a class of generalized friction models has been demonstrated in the framework of the quasi-1D approximation of thin elastic bodies in contact~\cite{Putelat2017}. Yet, the degree of generality of such slip pulses, and most importantly their dynamic stability and dimensionality dependence, remain rather poorly understood.

In this Letter, we establish a surprising connection between the two apparently disconnected classes of problems described above; we show that interfaces described by generic friction laws feature unstable steady-state slip pulses in the quasi-1D approximation. These unstable slip pulses of linear size $L^*$ and characteristic amplitude are hypothesized to serve as ``critical nuclei'' for the onset of rapid slip along quiescent interfaces, in a non-equilibrium analogy to equilibrium first-order phase transitions; that is, we propose an intimate relation between unstable slip pulses and earthquake nucleation.

These ideas are first quantitatively supported by dynamical quasi-1D calculations. Then the nucleation length $L^*$ is shown to exist also when 2D elastodynamics are considered, but that the existence of a fracture mechanics~\cite{freund1998dynamic,Andrews1976} Griffith-like length $L_G$ gives rise to sustained slip pulses. The analysis culminates in a 2D phase-diagram~\cite{Gabriel2012}, highlighting the roles of unstable slip pulses in earthquake nucleation as a non-equilibrium first-order phase transition.

{\em Spatially-extended frictional systems and generic friction laws.---} The dynamics of spatially-extended frictional systems emerge from the coupling between the bulk dynamics of the deformable bodies in contact and the frictional interaction of the bodies along the contact interface. Bulk dynamics are described by continuum momentum balance $\rho\ddot{\bm u}(\bm r, t)\=\nabla\!\cdot{\bm \sigma}(\bm r, t)$, where $\rho$ is the mass density, $\bm u$ and $\bm r$ are the $\dbar$-dimensional displacement and position vector fields respectively, $t$ is time and $\bm \sigma$ is the stress tensor field. $\bm \sigma$ is typically related to $\bm u$ through Hooke's law, resulting in bulk linear elastodynamics.

The interfacial constitutive (friction) law relates the slip velocity (the relative interfacial velocity), interfacial stresses and the structural state of the interface. In $\dbar\=2$, when sliding takes place along $x$ at $y\=0$, the slip velocity reads $v(x,t)\!\equiv\!\dot{u}_x(x,y\=0^+,t)\!-\!\dot{u}_x(x,y\=0^-,t)$ ($+/-$ correspond to the upper/lower bodies, respectively) and the friction stress reads $\tau(x,t)\!\equiv\!\sigma_{xy}(x,y\=0,t)$. The structural state of the interface is described by an internal-state field $\phi(x,t)$, which satisfies its own evolution equation. Finally, $\tau(v,\phi)\=\sigma\sgn(v)f(|v|,\phi)$, where $\sigma\!\equiv\!-\sigma_{yy}(x,y\=0,t)$ is the interfacial normal stress and $f(|v|,\phi)$ is the friction coefficient, and $\dot\phi\=g(|v|,\phi)$. This is the rate-and-state friction constitutive framework~\cite{Ruina1983,Marone1998a,Nakatani2001,Baumberger2006}.
\begin{figure}[ht]
 \centering
 \includegraphics[width=0.94\columnwidth]{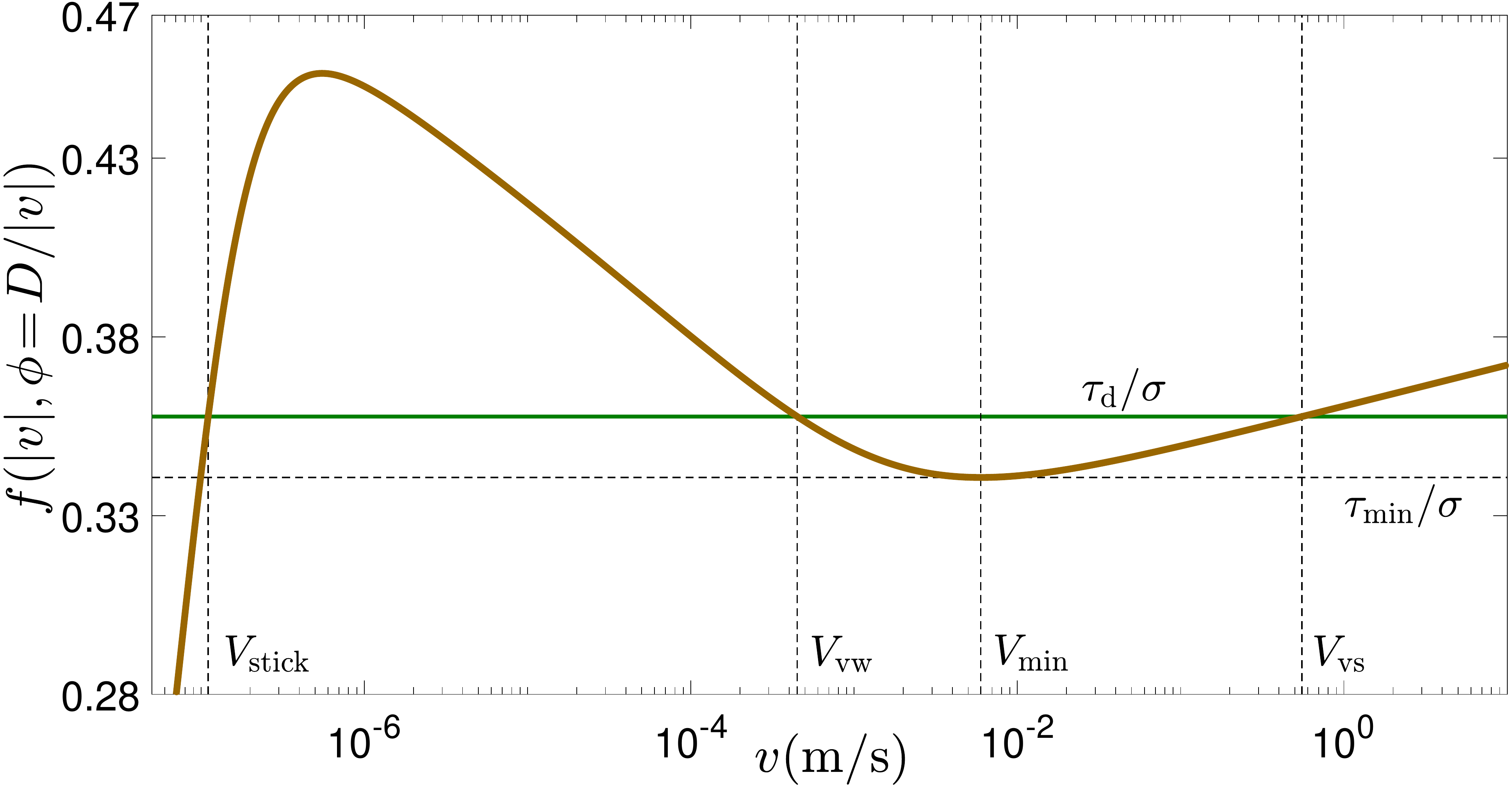}
 \caption{A generic N-shaped steady-state friction coefficient $f\!=\!\tau/\sigma$ vs.~slip velocity $v$, featuring a minimum at $(V_{\rm min}, \tau_{\rm min}/\sigma)$. The solid horizontal line is the driving shear stress $\tau_{\rm d}\!>\!\tau_{\rm min}$, intersecting the N-shaped\ friction law at $3$ velocities: $V_{\rm stick}\!\ll\!V_{\rm min}$ on the extremely low $v$ velocity-strengthening branch, $V_{\rm vw}\!<\!V_{\rm min}$ on the velocity-weakening branch and $V_{\rm vs}\!>\!V_{\rm min}$ on the high-$v$ velocity-strengthening branch. Note that the actual numbers used in the figure are characteristic of some laboratory experiments~\cite{SM}, but the results derived from them below are relevant to a broad range of materials and physical situations.}
 \label{fig:friction_law}
\end{figure}

We use constitutive functions $f(|v|,\phi)$ and $g(|v|,\phi)$ that capture the generic properties of frictional interfaces; first, we set $g(|v|,\phi)\!=\!1-|v|\phi/D$~\cite{Ruina1983,Baumberger2006,Bhattacharya2014,Marone1998a,Nakatani2001,SM}, where $\phi$ represents the typical age/maturity of contact asperities that compose the spatially-extended interface, such that $\phi\=t$ accounts for frictional aging/healing in the absence of slip, $v\=0$, and $\phi\=D/|v|$ accounts for frictional rejuvenation over characteristic slip $D$
in the presence of slip, $v\!\ne\!0$. Second, we use the function $f(|v|,\phi\=D/|v|)$~\cite{aldam2017a,SM} plotted in Fig.~\ref{fig:friction_law}; this N-shaped steady-state friction curve features a velocity-strengthening branch at extremely small $v$'s, essentially representing quiescent/locked interfacial states, a velocity-weakening branch at intermediate $v$'s and another velocity-strengthening branch beyond a high-$v$ minimum~\cite{Bar-Sinai2012,Bar-Sinai2014,Bar-Sinai2015a,SM}. This generic friction curve is supported by extensive experiments and theoretical considerations~\cite{Bar-Sinai2014}.

{\em The existence and properties of 1D steady-state pulses.---} The coupled interface-bulk problem defined above poses great mathematical challenges. To simplify things, we first consider two long and thin linear elastic bodies of height $H$ in frictional contact, such that $\rho\ddot{\bm u}\=\nabla\!\cdot{\bm \sigma}$ reduces to~\cite{Bar-Sinai2012,Bar-Sinai2013}
\begin{equation}
\label{eq:FB}
H\Gb\left(c^{-2}\partial_{tt}-\partial_{xx}\right)u\!\left(x,t\right)=\tau_{\rm d}-\tau\left[v\!\left(x,t\right),\phi\!\left(x,t\right)\right]\ ,
\end{equation}
where $u\!\equiv\!u_x$, $\Gb$ and $c$ are the effective shear modulus and wave-speed~\cite{Bar-Sinai2012,Bar-Sinai2013}, respectively, and $\tau_{\rm d}$ is a constant driving stress (see Fig.~\ref{fig:friction_law}). Quasi-1D traveling steady-state solutions then satisfy~\cite{SM}
\begin{eqnarray}
\label{eq:SS_FB}
&\Gb H c^{-1}(1-\beta^2)\beta^{-1}\,dv(\xi)/d\xi=\tau_{\rm d}-\tau (v(\xi ),\phi (\xi ))\ ,\\
\label{eq:SS_phi}
&\beta\,c\,d\phi(\xi)/d\xi=\phi(\xi)v(\xi)/D-1\ ,
\end{eqnarray}
where we defined a co-moving coordinate $\xi\!\equiv\!x\!-\!\beta c\,t$, integrated out $u$ and eliminated partial time-derivatives.
\begin{figure}[ht]
 \centering
 \includegraphics[width=0.92\columnwidth]{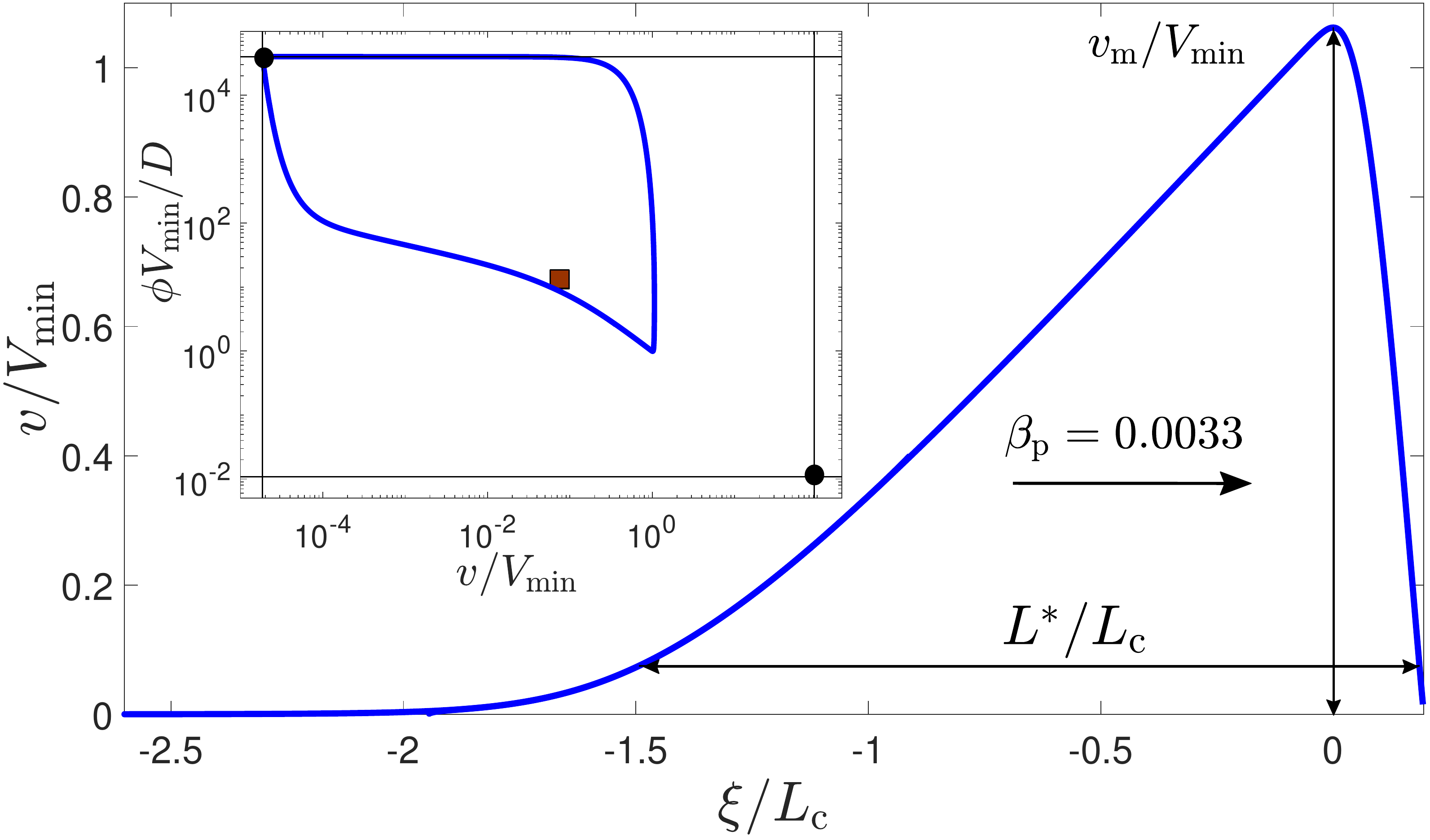}
 \caption{A slip pulse solution corresponding to $\tau_{\rm d}$ of Fig.~\ref{fig:friction_law}, featuring a typical width $L^*$ (see text for exact definition) and a maximal slip velocity $v_{\rm m}$~\cite{SM}. Length is measured in units of the velocity-weakening nucleation length $L_{\rm c}$~\cite{Bar-Sinai2013,aldam2017a,SM} and velocity in units of $V_{\rm min}$. (inset) The solution in the $\phi\!-\!v$ plane (the black circles correspond to the $V_{\rm stick}$ and $V_{\rm vs}$ fixed-points and the brown square to the $V_{\rm vw}$ fixed-point, see Fig.~\ref{fig:friction_law}).}
 \label{fig:pulse}
\end{figure}

\begin{figure}[ht]
 \centering
 \includegraphics[width=\columnwidth]{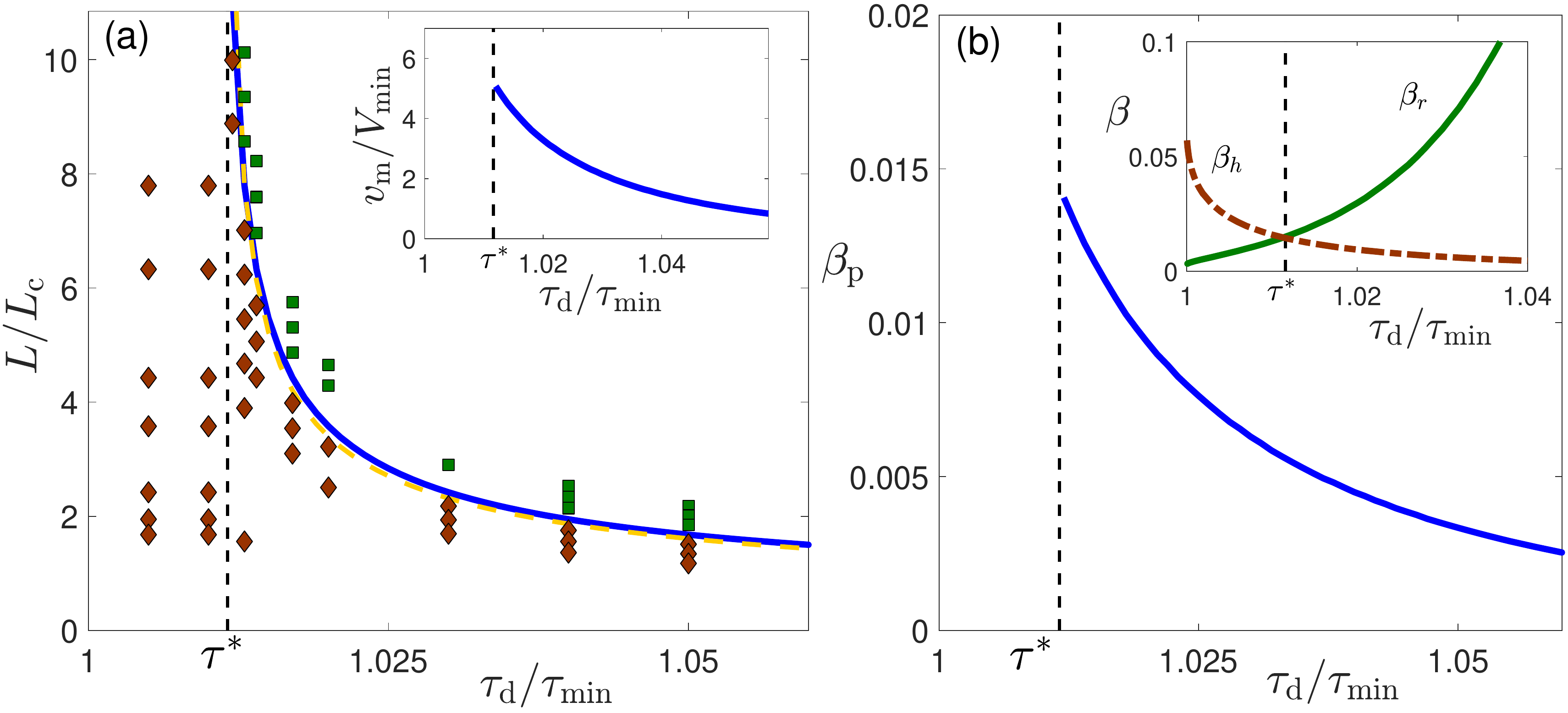}
 \caption{(a) The normalized perturbation width $L/L_{\rm c}$ vs.~$\tau_{\rm d}/\tau_{\rm min}$. The theoretical prediction $L^*$ (solid blue line) separates dynamic perturbations that lead to nucleation (green squares) from those that decay (brown diamonds), see text for details, and closely follows the theoretical prediction $\sim\!1/\sqrt{\tau_{\rm d}\!-\!\tau^*}$ (dashed yellow line)~\cite{SM}, where $\tau^*$ is marked by the vertical dashed line. (inset) $v_{\rm m}/V_{\rm min}$ vs.~$\tau_{\rm d}/\tau_{\rm min}$. (b) The dimensionless pulse propagation velocity $\beta_{\rm p}$ vs.~$\tau_{\rm d}/\tau_{\rm min}$. (inset) The dimensionless front propagation velocity $\beta$ (solid green line for rupture fronts, $\beta_{\rm r}$, and dashed-dotted brown line for healing fronts, $\beta_{\rm h}$) vs.~$\tau_{\rm d}/\tau_{\rm min}$. The two curves intersect at $\tau^*$.}
 \label{fig:pulse_props}
\end{figure}

Steady-state pulses, featuring a steadily travelling slipping region (cf.~Fig.~\ref{fig:pulse}), can be thought of as composed of interacting rupture and healing fronts that propagate at the same velocity. Such fronts connect velocity-strengthening (i.e.~stable) solutions of $\tau(|v|,\phi\=D/|v|)\=\tau_{\rm d}$, see Fig.~\ref{fig:friction_law}. In particular, in steady-state rupture fronts the homogeneous $V_{\rm vs}$ state invades the homogeneous $V_{\rm stick}$ state~\cite{Rubino2000,Bizzarri2003,Ben-David2010a,Bar-Sinai2012,Svetlizky2014,Putelat2017}, both defined in Fig.~\ref{fig:friction_law}, and vice versa for steady-state healing fronts. We found these solutions and calculated their dimensionless propagation velocity $\beta_{\rm r,h}(\tau_{\rm d})$ (for rupture/healing fronts, respectively), as shown in the inset of Fig.~\ref{fig:pulse_props}b. The two functions exhibit opposite trends and intersect at $\tau^*$~\cite{Putelat2017}.

At $\tau_{\rm d}\=\tau^*$, rupture and healing fronts propagate at the same velocity, $\beta_{\rm r}(\tau^*)\=\beta_{\rm h}(\tau^*)$, and hence can be superimposed without interaction to form a pulse of infinite width. As $\tau_{\rm d}$ is increased above $\tau^*$, the two fronts interact, leading to pulses of finite width $L^*(\tau_{\rm d})$ and propagation velocity $\beta_{\rm p}(\tau_{\rm d})$. The existence of such pulses is explicitly demonstrated in Fig.~\ref{fig:pulse}. In Fig.~\ref{fig:pulse_props}a we show the pulse width $L^*(\tau_{\rm d})$, defined as the distance between the points at which the slip velocity drops to $V_{\rm vw}$ (cf.~Fig.~\ref{fig:friction_law}). A scaling theory predicts that $L^*/L_{\rm c}$ diverges as $\tau^*/\sqrt{\tau_{\rm d}\!-\!\tau^*}$ for $\tau_{\rm d}\!\to\!\tau^*$~\cite{SM}, which is shown to quantitatively agree with the numerical results (dashed yellow line in Fig.~\ref{fig:pulse_props}a). The maximal slip velocity $v_{\rm m}(\tau_{\rm d})$ (see definition in Fig.~\ref{fig:pulse}), plotted in the inset of Fig.~\ref{fig:pulse_props}a, and the propagation velocity of slip pulses $\beta_{\rm p}(\tau_{\rm d})$, plotted in Fig.~\ref{fig:pulse_props}b, also increase with decreasing $\tau_{\rm d}$.

Finally, we note that while it is physically intuitive and appealing to think of slip pulses as interacting rupture and healing fronts, the existence of the latter is not a necessary condition for the existence of the former. That is, slip pulses exist also for a steady-state friction laws that do not feature a minimum, for which steady-state fronts --- and consequently a finite $\tau^*$ at which $L^*$ diverges --- do not exist~\cite{SM}.

\begin{figure*}[ht]
 \centering
 \includegraphics[width=0.7\textwidth]{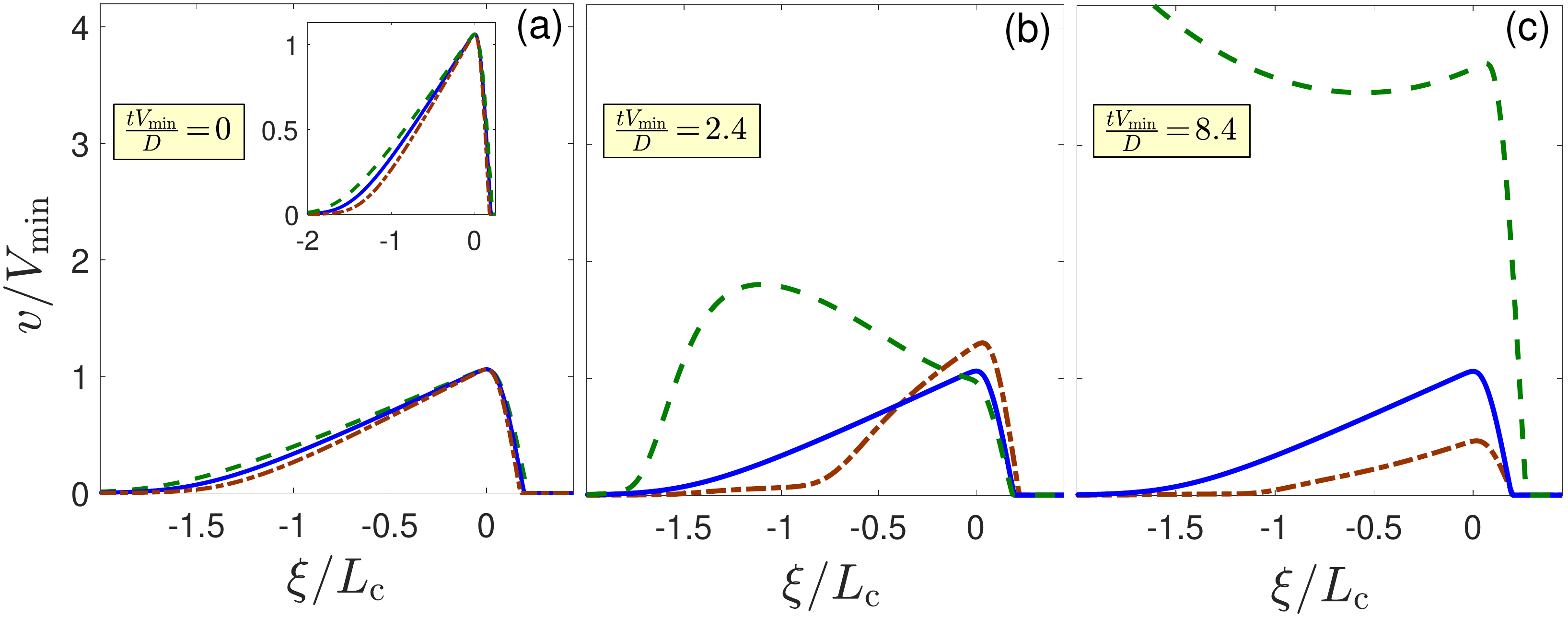}
 \caption{(a) A steady-state pulse (solid blue line), which is slightly stretched (dashed green line), $L\!>\!L^*$, and compressed (dashed-dotted brown line), $L\!<\!L^*$, at $t\!=\!0$. (inset) Zoom in. As time progresses, (b) and (c), the stretched perturbation grows and expands, while the compressed one decays.}
 \label{fig:pulse_stability}
\end{figure*}

\emph{Unstable pulses as critical nuclei in a non-equilibrium first-order phase transition.---} We next perform numerical stability analysis of slip pulses. That is, we use steady-state pulse solutions as initial conditions and perturb them by slightly stretching/compressing the co-moving coordinate $\xi$, see Fig.~\ref{fig:pulse_stability}a. Here steady-state conditions are not enforced, i.e.~we transform the equations to the co-moving frame of reference {\em without} eliminating partial time-derivatives~\cite{SM}, and track the time evolution of perturbations, see Fig.~\ref{fig:pulse_stability}b-c. It is observed that the perturbation with $L\!>\!L^*$ grows in amplitude and expands in size, while the perturbation with $L\!<\!L^*$ decays. That is, 1D steady-state pulses are intrinsically unstable.

The unstable nature of 1D steady-state slip pulses may imply that they play no role in fault dynamics. We propose, instead, that in fact they may serve as ``critical nuclei'' for the transition from an almost non-slipping state $V_{\rm stick}$ to a slipping state $V_{\rm vs}$ in a non-equilibrium analogy to equilibrium first-order phase transitions~\cite{stanley1971phase}. That is, we propose that unstable slip pulses provide a dynamical mechanism for the nucleation of rapid slip along frictional interfaces that are initially at (or nearly at) rest, a regime that is not commonly studied in the literature. To test this idea, we introduced perturbations as initial conditions in the dynamical equations obtained by stretching ($L\!>\!L^*(\tau_{\rm d})$) or compressing ($L\!<\!L^*(\tau_{\rm d})$) the steady-state pulse solutions corresponding to $L^*(\tau_{\rm d})$ for each $\tau_d\!>\!\tau^*$, and by solving for a reduced $\tau_d$ using steady-state pulse solutions corresponding to $L\=L^*$ for each $\tau_d\!<\!\tau^*$. We then tracked the system's evolution to determine whether the perturbations decay back to $V_{\rm stick}$ or bring the system to $V_{\rm vs}$.

The results, over a range of $L$ and $\tau_{\rm d}$ values, are superimposed on Fig.~\ref{fig:pulse_props}a. The figure provides compelling evidence that the theoretical prediction $L^*(\tau_{\rm d})$ indeed quantitatively predicts the fate of dynamic perturbations, i.e.~perturbations with $L\!<\!L^*$ or $\tau_d\!<\!\tau^*$ (brown diamonds) decay back to $V_{\rm stick}$ and those with $L\!>\!L^*$ (green squares) grow and bring the system to $V_{\rm vs}$, lending strong support to the proposed connection between unstable slip pulses and earthquake nucleation.
\begin{figure}[ht]
 \centering
 \includegraphics[width=0.98\columnwidth]{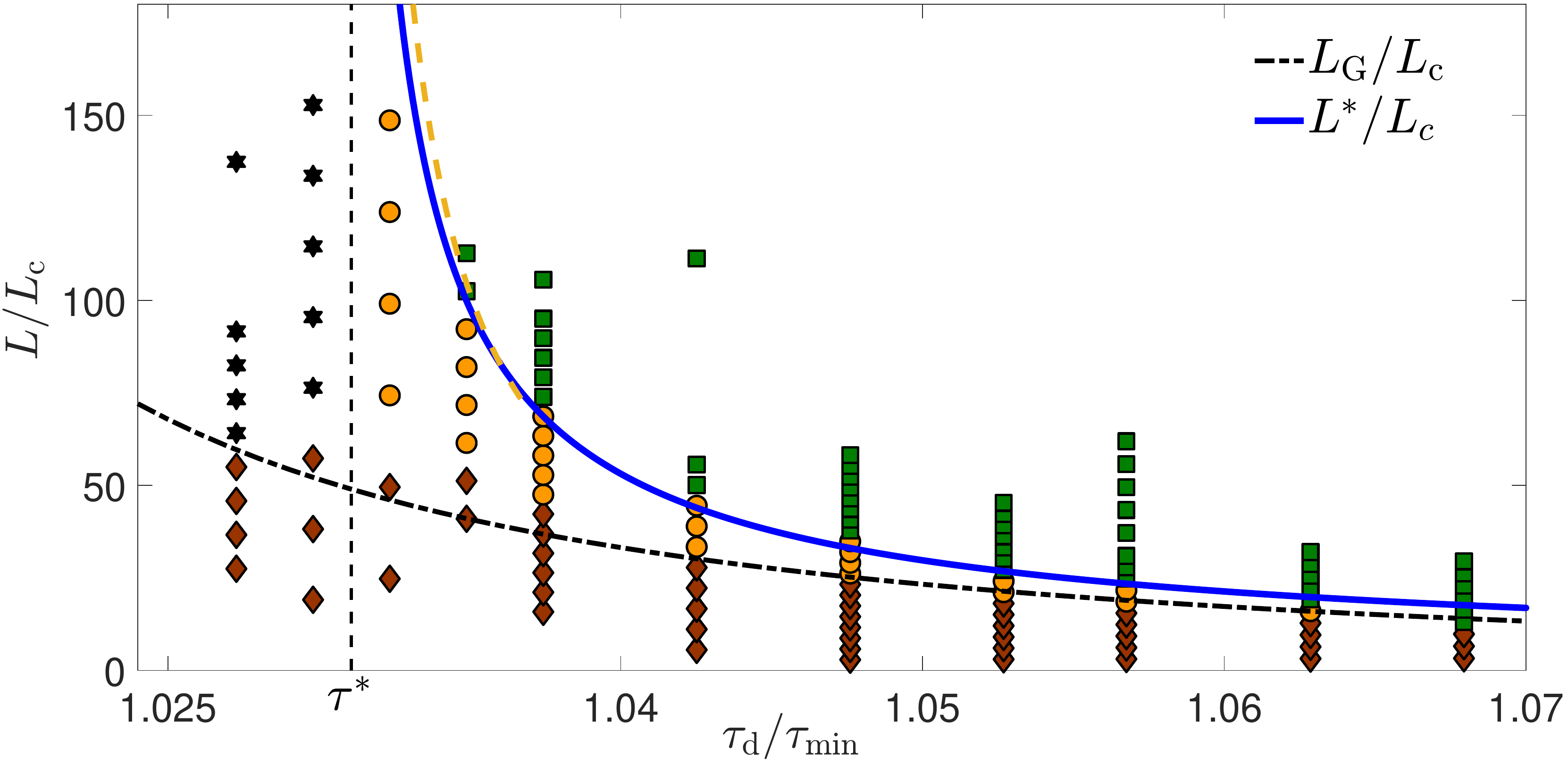}
 \caption{A 2D phase-diagram, the counterpart of the 1D phase-diagram of Fig.~\ref{fig:pulse_props}a. Dynamic perturbations that decay without propagation (brown diamonds) appear below the Griffith-like length $L_{\rm G}(\tau_{\rm d})$ (dashed-dotted black line, obtained analytically, see text for details). Dynamic perturbations that lead to nucleation (green squares) appear above $L^*(\tau_{\rm d})$ (solid blue line, estimated numerically), which follows the theoretical prediction $\sim\!1/(\tau_{\rm d}-\tau^*\!)$ (dashed yellow line) when $\tau_{\rm d}$ is close to $\tau^*$ (marked by the vertical dashed line)~\cite{SM}. For $L\!>\!L_{\rm G}$ and $\tau_{\rm d}\!<\!\tau^*$, dynamical perturbations decay with some propagation of transient pulses (black hexagrams) and for $L_{\rm G}\!<\!L\!<\!L^*$ and $\tau_{\rm d}\!>\!\tau^*$, sustained pulses exist (orange circles, see text for details). Movies are available at~\cite{SM}.}
 \label{fig:phase-diagram}
\end{figure}

\emph{2D phase-diagram: The Griffith-like length and sustained pulses.---} The concepts and physical picture developed above are expected to be $\dbar$-independent, and hence we expect the nucleation length $L^*(\tau_{\rm d})$ to exist also in $\dbar\!>\!1$. For $\dbar\!>\!1$, fronts and pulses are accompanied by a crack-like singularity near their edges~\cite{Ben-Zion2001,Ben-Zion2008,Bizzarri2003,Svetlizky2014,SM}, associated with a finite energy flux that is required to balance near-edge frictional dissipation per unit area, $G_{\rm c}$ (an effective fracture energy)~\cite{Brener2002,Svetlizky2014,Kammer2015,Bayart2015,Svetlizky2017,SM}. Consequently, for a given $G_{\rm c}$, there exists a Griffith-like length $L_{\rm G}(\tau_{\rm d})\=4\mu\pi^{-1}G_{\rm c}(\tau_{\rm d}-\tau_{\rm res})^{-2}$~\cite{freund1998dynamic,Andrews1976}, where $\mu$ is the shear modulus and $\tau_{\rm res}$ is the residual shear stress left behind the edge, below which no front/pulse propagation is possible.

Using $L^*(\tau_{\rm d})$ and $L_{\rm G}(\tau_{\rm d})$, we can predict the salient features and topology of the $L\!-\!\tau_{\rm d}$ phase-diagram for $\dbar\!>\!1$. First, $L^*/L_{\rm c}$ is predicted to diverge at a finite $\tau^*$ as $\tau^*/(\tau_{\rm d}-\tau^*)$ for $\dbar\!>\!1$~\cite{SM}. As $L_{\rm G}(\tau_{\rm d})\!\sim\!(\tau_{\rm d}-\tau_{\rm res})^{-2}$ is a minimal condition for front/pulse propagation, we expect $L_{\rm G}\!<\!L^*$ and $\tau_{\rm res}\!<\!\tau^*$. Consequently, for $L\!<\!L_{\rm G}$, we expect perturbations to decay without propagation, and hence no nucleation to occur, simply because no front/pulse can propagate. For $L\!>\!L^*$, we expect perturbations (of sufficiently large amplitude) to lead to the nucleation of the $V_{\rm vs}$ phase through propagating rupture fronts. For $L\!>\!L_{\rm G}$ and $\tau_{\rm d}\!<\!\tau^*$, we expect no nucleation to occur, but that the decay of perturbations to be different from that in the regime $L\!<\!L_{\rm G}$ and involve front/pulse propagation. Finally, for perturbations with $L_{\rm G}\!<\!L\!<\!L^*$ under $\tau_{\rm d}\!>\!\tau^*$, new dynamical modes that have no 1D analog, might emerge.

To test these predictions, we performed spectral boundary integral method~\cite{Geubelle1995,Morrissey1997,Breitenfeld1998} calculations for infinite $\dbar\=2$ systems under anti-plane shear (mode-III)~\cite{freund1998dynamic}. The basic field in this problem, $u_z(x,y,t)$ ($z\,\bot\,x,y$), satisfies the bulk elastodynamic equation $\mu\nabla^2 u_z\=\rho\ddot{u}_z$, together with $v(x,t)\!\equiv\!\dot{u}_z(x,y\=0^+,t)\!-\!\dot{u}_z(x,y\=0^-,t)$ and $\tau(x,t)\!\equiv\!\sigma_{yz}(x,y\=0,t)\=\mu\,\partial_yu_z(x,y\=0,t)$. Furthermore, to test the robustness of the emerging physical picture for different types of initial perturbations, we consider here Gaussian perturbations (the perturbation's width $L$ is defined as $10$ Gaussian standard deviations~\cite{SM}), which are somewhat more generic. The results are presented in Fig.~\ref{fig:phase-diagram}, where the theoretical prediction for $L_{\rm G}(\tau_{\rm d})$ is added (dashed-dotted black line, details about the estimation of $G_{\rm c}$ and $\tau_{\rm res}$ can be found in~\cite{SM}). First, we observe that $L_{\rm G}(\tau_{\rm d})$ quantitatively predicts the boundary below which perturbations decay without propagation. Second, we observe that there exists a vertical boundary (dashed line), which is interpreted as $\tau^*$, such that no nucleation occurs for $\tau_{\rm d}\!<\!\tau^*$, yet the decay for $L\!>\!L_{\rm G}$ involves propagation of transient pulses, as predicted theoretically. Third, there exists a phase boundary (solid blue line), which appears to diverge at $\tau^*$ and hence interpreted as $L^*(\tau_{\rm d})$, above which nucleation occurs through rupture front propagation. The numerical $L^*/L_{\rm c}$ line is consistent with the theoretical prediction (cf.~dashed yellow line in Fig.~\ref{fig:phase-diagram}). Movies are available at~\cite{SM}.

Finally, for $L_{\rm G}\!<\!L\!<\!L^*$ and $\tau_{\rm d}\!>\!\tau^*$, sustained pulses that do not appear to exist in 1D emerge~\cite{Perrin1995, Zheng1998}. In this dynamical regime, a pair of pulses move away from one another, apparently indefinitely (see movie at~\cite{SM}). While these pulses do not strictly reach steady-state conditions for computationally feasible system sizes~\cite{SM}, it is clear that they leave behind them a $V_{\rm stick}$ state and hence they do not lead to the nucleation of the $V_{\rm vs}$ phase. The results presented in Fig.~\ref{fig:phase-diagram} appear to be independent of the amplitude of perturbations, as long as it is larger than $V_{\rm vw}$~\cite{SM}. Note that while in $\dbar\=1$ no edge singularity exists, a Griffith-like length (which scales as $\sim\!\sqrt{\Gb\,H\,G_c}\,(\tau_{\rm d}-\tau_{\rm res})^{-1}$) can still be defined using global energy balance considerations~\cite{Puzrin2005,SM}. Yet, we found no trace for this length in our 1D phase-diagram in Fig.~\ref{fig:pulse_props}a.

The phase-diagram in Fig.~\ref{fig:phase-diagram} may appear somewhat reminiscent of the computational results of~\cite{Gabriel2012}, obtained in a large parametric study of in-plane (mode-II) dynamic rupture styles of faults featuring a finite shear strength and strong velocity-weakening friction. Yet, there are important differences between the two works. Most notably, we provide here a theoretical understanding of the phase-boundaries $L^*(\tau_{\rm d})$ (associated with critical nuclei) and $L_G(\tau_{\rm d})$ (associated with a Griffith-like length), which is not developed in~\cite{Gabriel2012}, and we directly relate $L^*(\tau_{\rm d})$ in 1D to steady-state slip pulses and their stability, which are not discussed in~\cite{Gabriel2012}.

\emph{Concluding remarks and prospects.---} We developed a comprehensive physical picture of rapid slip nucleation along quiescent frictional interfaces, highlighting the role of unstable slip pulses as critical nuclei of size $L^*$ in a non-equilibrium analogy to equilibrium first-order phase transitions. We also elucidated the conditions for the emergence of various propagative slippage modes (rupture styles), including rupture fronts, decaying pulses, transient pulses and sustained pulses~\cite{Zheng1998, Gabriel2012}. We stress that the physics behind the nucleation length $L^*$, associated with abrupt and stochastic processes, is qualitatively different from that of $L_{\rm c}$, which is intrinsically related to a deterministic velocity-weakening linear frictional instability typically associated with precursory slow slip. Seismological evidence for such qualitatively different nucleation dynamics has been recently discussed~\cite{Gomberg2018} and should be further explored in the future.

\emph{Acknowledgments.---} E.B.~and J.-F.M.~acknowledge support from the Rothschild Caesarea Foundation. E.B.~acknowledges support from the Israel Science Foundation (Grant No.~295/16). J.-F.M.~and F.B.~acknowledge support from the Swiss National Science Foundation (Grant No.~162569). M.A.~thanks Y.~Lubomirsky for assistance in numerical steady-state calculations in 1D. The authors thank E.~Dunham and R.~Viesca for useful discussions.

%

\clearpage

\onecolumngrid
\begin{center}
	\textbf{\large Supplemental Material for: ``Unstable slip pulses and\\ earthquake nucleation as a non-equilibrium first-order phase-transition''}
\end{center}

\setcounter{equation}{0}
\setcounter{figure}{0}
\setcounter{section}{0}
\setcounter{table}{0}
\setcounter{page}{1}
\makeatletter
\renewcommand{\theequation}{S\arabic{equation}}
\renewcommand{\thefigure}{S\arabic{figure}}
\renewcommand{\thesection}{S-\Roman{section}}
\renewcommand*{\thepage}{S\arabic{page}}
\renewcommand{\bibnumfmt}[1]{[S#1]}
\renewcommand{\citenumfont}[1]{S#1}

\twocolumngrid
The goal of this document is to provide additional technical details regarding the results reported on in the manuscript.

\subsection{The friction law}
\label{sec:fric}

The friction law used in this work, and whose steady-state behavior is plotted in Fig.~1 in the manuscript, is the same one used previously in~\cite{SMAldam2017a}. The friction law is defined by the relation between the shear stress $\tau\!\equiv\!\sigma_{xy}$ and the compressive normal stress $\sigma\!\equiv\!-\sigma_{yy}$ at the interface,  $\tau\=\sigma \sgn(v)f\left(\left|v\right|,\phi\right)$, and by the evolution equation for state variable $\phi$, $\dot\phi\=g\left(\left|v\right|,\phi\right)$. The constitutive functions $f\left(\left|v\right|,\phi\right)$ and $g\left(\left|v\right|,\phi\right)$ used in this work take the form
\begin{eqnarray}
\label{eq:f}
&& f\left(\left|v\right|,\phi\right)=\left[1+b \log \left(1+\frac{\phi }{\phi _*}\right)\right] \times \\
&&\qquad\qquad\qquad\qquad \left[\frac{f_0}{\sqrt{1+\left(v_*/v\right)^2}}+\alpha  \log \left(1+\frac{\left| v\right| }{v_*}\right)\right]\ ,\nonumber\\
\label{eq:g}
&&g\left(\left|v\right|,\phi\right)=1-\frac{\left| v\right|\phi}{D}\sqrt{1+\left(v_*/v\right)^2}\ ,
\end{eqnarray}
where $\phi$ represents the typical age/maturity of contact asperities that compose the spatially-extended interface~\cite{SMBaumberger2006,SMPutelat2011}.

Equation~\eqref{eq:f} identifies (up to $\log^2$ terms) with the conventional Ruina-Rice rate-and-state friction~\cite{SMRuina1983,SMMarone1998,SMMarone1998a,SMNakatani2001,SMBhattacharya2014}, $f\!\simeq\!f_0+f_0b \log(\phi/\phi_*)+\alpha \log(\left|v\right|/v_*)$, if ``1'' in both $\log$ terms in is omitted and $\sqrt{1+\left(v_*/v\right)^2}\!\to\!1$. Here $f_0$ sets the scale of the dimensionless frictional resistance (friction
coefficient), $b$ is the aging coefficient and $\alpha$ is related to the thermally-activated rheology of contact asperities~\cite{SMBaumberger2006,SMPutelat2011}. In many cases, we have $f_0b\!>\!\alpha$, which implies that steady-state friction (where $\phi\=D/|v|$) is velocity-weakening, i.e. $f$ decreases with increasing $|v|$. The constitutive functions $f\left(\left|v\right|,\phi\right)$ and $g\left(\left|v\right|,\phi\right)$ in Eqs.~\eqref{eq:f}-\eqref{eq:g} go beyond the conventional rate-and-state friction relation in three respects; first, the ``1'' in the $\alpha$ term and the function $1/\sqrt{1+\left(v_*/v\right)^2}$ that multiplies $f_0$ (both with a very small $v_*$) that ensure both that $f$ vanishes as $v\!\to\!0$ and that a low-$v$ velocity-strengthening regime exists prior to the velocity-weakening regime~\cite{SMEstrin1996}. Second, the ``1'' in the $b$ term that implies that $\phi_*$ is a short-time cutoff on logarithmic aging~\cite{SMMarone1998,SMNakatani2006,SMBen-David2010}, which in turn leads to a high-$v$ minimum in the steady-state $f$ and to another velocity-strengthening regime, as documented for many materials~\cite{SMBar-Sinai2014}. Third, the function $\sqrt{1+\left(v_*/v\right)^2}$ that appears also in $g\left(\left|v\right|,\phi\right)$, which ensures that for vanishingly small steady-state velocities, $\phi$ saturates after extremely long times to a finite value of $D/v_*$, rather than diverges. Note that the authors of~\cite{SMPerrin1995} invoked a similar regularization of the $\dot\phi$ equation; however, while
this regularization is crucial for the existence of their 2D steady-state pulse solutions, in our case it makes no difference (even quantitatively the differences are minute) because the steady frictional resistance increases smoothly from zero at vanishing steady-state sliding velocities, hence in the main text we omitted $\sqrt{1+\left(v_*/v\right)^2}$ from $g\left(\left|v\right|,\phi\right)$ (though it is included in the calculations).

\subsection{The calculation of the velocity-weakening linear frictional instability nucleation length \texorpdfstring{$L_{\rm c}$}{the velocity-weakening nucleation length}}

The calculation of the nucleation length $L_{\rm c}$, associated with the velocity-weakening linear frictional instability, appeared in previous publications (e.g.~\cite{SMAldam2017a}). The main procedure is repeated here for completeness. We perform a linear stability analysis of an interface  steadily sliding at a velocity $V$. We start with the steady-state frictional relation $\tau=\sigma{f_{\rm ss}(V)}$, where $f_{\rm ss}$ is plotted in Fig.~1 of the manuscript, from which we obtain
\begin{equation}
\delta\tau=\sigma\delta{f}+f\delta\sigma=\sigma\left(f_v\delta{v}+f_\phi\delta\phi\right)+f\delta\sigma\ ,
\end{equation}
where we use the shorthand $f_v\=\partial_vf$, evaluated at steady-state, and similar notations used for $\phi$ derivatives and later on for derivatives of $g$. The normal stress at the interface, $\sigma$, is assumed to be constant (i.e.~physical mechanisms that generate normal stress variations, such as bi-material contrast~\cite{SMAldam2016,SMAldam2017a}, are not considered here). In 1D, i.e.~in the small height limit, $\sigma$ at the interface equals, by definition, to $-\sigma_{yy}(y\=H)$~\cite{SMBar-Sinai2013}. As $\sigma$ is constant, $\delta\sigma\=0$ and the last term in the above equation can be dropped. We assume that all of the fields are proportional to a Fourier mode $e^{\Lambda{t}-i k x}$ such that $\delta{v}\=\Lambda\delta{u}$ and $\delta \phi\=\frac{g_v}{\Lambda -g_\phi}\delta{v}$. Putting it all together, we find
\begin{equation}
\label{eq:spec}
\delta\tau=\sigma\left(f_v+f_\phi\frac{g_v}{\Lambda -g_\phi}\right)\Lambda\delta{u}\ .
\end{equation}
$\delta\tau$ is obtained from the bulk solution~\cite{SMAldam2016,SMAldam2017a}. In this work the bulk is described by linear elasticity~\cite{SMLandau1986} and for the calculation of $L_c$ we focus on the quasi-static limit, which implies that $\delta\tau$ does not depend on $\Lambda$. At the critical wavenumber $k_c\=2\pi/L_{\rm c}$, we have $\Lambda\=i\omega$, and we can decompose Eq.~\eqref{eq:spec} into its real and imaginary parts
\begin{equation}
G\left(\frac{2 \pi }{L_{\rm c}}\right)+\frac{f_\phi g_v \sigma  \omega ^2}{g_\phi^2+\omega ^2}=0\ ,\qquad f_v \sigma  \omega -\frac{f_\phi g_v g_\phi \sigma  \omega }{g_\phi^2+\omega ^2}=0\ ,
\end{equation}
where we defined the elastic transfer function $G(k)\!\equiv\!\delta\tau/\delta{u}$. The imaginary part implies
\begin{equation}
\omega^2=\frac{g_\phi (f_\phi g_v-f_v g_\phi)}{f_v}=-\frac{g_{\phi }^2}{f_v}f_{\rm ss}'(V)\ ,
\end{equation}
which means there is a solution only for $f_{\rm ss}'(V)<0$, i.e.~for velocity-weakening friction (note that $f_v\!>\!0$). Substituting this result into the real part above, we obtain
\begin{equation}
G\left(\frac{2 \pi }{L_{\rm c}}\right)=\sigma  g_{\phi } f_{\rm ss}'(V)\ ,
\end{equation}
from which $L_{\rm c}$ can be extracted. Note that $g_{\phi}\!<\!0$.

In the 1D approximation we have $G(k)=\Gb H k^2$~\cite{SMBar-Sinai2013}, which implies
\begin{equation}
L_{\rm c}^{\left(\rm 1D\right)}=2 \pi  \sqrt{\frac{\Gb H}{\sigma  g_{\phi } f_{\rm ss}'(V)}}\ .
\label{eq:Lc1D}
\end{equation}
$L_{\rm c}^{\left(\rm 1D\right)}$ (without the superscript) is used to normalize quantities of length dimension in the 1D part of the manuscript. In 2D, under mode-III symmetry conditions, we have $G(k)=\mu\left|k\right|/2$, which implies
\begin{equation}
L_{\rm c}^{\left(\rm 2D\right)}=\frac{\pi  \mu }{\sigma  g_{\phi } f_{\rm ss}'(V)}\ .
\label{eq:Lc2D}
\end{equation}
$L_{\rm c}^{\left(\rm 2D\right)}$ (without the superscript) is used to normalize quantities of length dimension in the 2D part of the manuscript. Finally, note that for $g(|v|,\phi)$ of Eq.~\eqref{eq:g}, we have $g_\phi\=-V/D$ in Eqs.~\eqref{eq:Lc1D}-\eqref{eq:Lc2D} (where $V\!\gg\!v_*$ is assumed).

\subsection{Scaling theory of $L^*(\tau_{\rm d})$ in 1D and 2D}

A simple scaling estimate of the steady-state Eq.~(2) in the manuscript reads
\begin{equation}
\frac{\Gb\,H}{\beta\,c} \frac{v}{L^*} \sim \tau_{\rm d} - \tau^* \ ,
\label{eq:momentum1d}
\end{equation}
where $\tau^*$ is used as the characteristic stress scale and $L^*$ as the characteristic length. A simple scaling estimate of the steady-state Eq.~(3) in the manuscript reads
\begin{equation}
\frac{\beta\,c\,\phi}{L^*} \sim \frac{\phi\,v}{D} \ ,
\label{eq:phi}
\end{equation}
where we are interested in the dynamic regime where the aging contribution (the ``1'' on the RHS) is negligible. Note that $\phi$ can be eliminated from both sides. Multiplying Eqs.~\eqref{eq:momentum1d}-\eqref{eq:phi}, we obtain
\begin{equation}
L^* \sim \sqrt{\frac{\Gb\,H\,D}{\tau_{\rm d} - \tau^*}} \ ,
\label{eq:L1d}
\end{equation}
which reveals both the scaling of $L^*$ as $\tau_{\rm d}\!\to\!\tau^*$ and its dependence on the bulk and interfacial properties.

In the manuscript we use the velocity-weakening nucleation length $L_c$ discussed above to normalize $L^*$. Using $L_{\rm c}^{\left(\rm 1D\right)}$ of Eq.~\eqref{eq:Lc1D} we obtain
\begin{equation}
\frac{L^*(\tau_{\rm d})}{L_{\rm c}^{\left(\rm 1D\right)}} = A_{\rm 1D}\sqrt{\frac{\tau^*}{\tau_{\rm d}-\tau^*}} \ ,
\label{eq:L*1d_norm}
\end{equation}
which is supported by the numerical results in the manuscript, cf.~the dashed yellow line in Fig.~3a in which $\tau^*$ is explicitly marked and $A_{\rm 1D}\=0.314$ is used.

To generate the corresponding scaling estimate in 2D, we need to replace Eq.~(2) in the manuscript by its 2D counterpart. The latter is given by the steady-state version of Eq.~\eqref{spectral} below and its scaling estimate reads
\begin{equation}
\frac{v}{\beta\,c_s} \sim \frac{\tau_{\rm d} - \tau^*}{\mu} \ ,
\label{eq:momentum2d}
\end{equation}
where again $\tau^*$ is used as the characteristic stress scale. Together with Eq.~\eqref{eq:phi} (where $c$ is replaced by $c_s$), which is independent of the spatial dimension, we obtain
\begin{equation}
L^* \sim \frac{\mu\,D}{\tau_{\rm d} - \tau^*} \ .
\label{eq:L2d}
\end{equation}
In fact, this relation can be obtained from the 1D result in Eq.~\eqref{eq:L1d} without any explicit knowledge of the 2D problem. To see this, note that 1D result in Eq.~\eqref{eq:L1d} should cross over to the 2D result for a typical system height $H$ that satisfies $L\!\sim\!H$ in both 1D and 2D, directly leading to Eq.~\eqref{eq:L2d}. Finally, normalizing by $L_{\rm c}^{\left(\rm 2D\right)}$ of Eq.~\eqref{eq:Lc2D}, we obtain
\begin{equation}
\frac{L^*(\tau_{\rm d})}{L_{\rm c}^{\left(\rm 2D\right)}} = A_{\rm 2D}\frac{\tau^*}{\tau_{\rm d}-\tau^*} \ ,
\label{eq:L*2d_norm}
\end{equation}
which is in reasonably good agreement with the numerical results in the manuscript, cf.~the dashed yellow line in Fig.~5 in which $\tau^*$ is explicitly marked and $A_{\rm 2D}\=0.430$ is used. Finally, note that the result in Eq.~\eqref{eq:L*2d_norm} is in fact valid also in 3D.

\subsection{Existence and stability of steady-state travelling solutions in the 1D approximation}
\label{sec:SS}

In the 1D limit, our equations take the form~\cite{SMBar-Sinai2012,SMBar-Sinai2013,SMPutelat2017}
\begin{align}
\label{eq:v}
&\partial_t{u}(x,t)=v(x,t) \ ,\\
\label{eq:D_EOM}
 & H \Gb\left(c^{-2}\partial_{tt}-\partial_{xx}\right)u(x,t)=\tau_{\rm d}-\sigma{f}\left[v(x,t),\phi(x,t)\right]\ , \\
\label{eq:D_phi}
&\partial_t\phi(x,t) =g\left[v(x,t),\phi(x,t)\right]\ .
\end{align}
A transformation to a co-moving frame of reference, moving at velocity $\beta{c}$, implies $\partial_x\!\to\!\partial_\xi$ and $\partial_{t}\!\to\!\partial_t\!-\!\beta{c}\partial_\xi$, where $\xi\!\equiv\!x\!-\!\beta{c}t$, leading to the following set of equations
\begin{eqnarray}
\label{eq:v_co}
&&\left(\partial_t-\beta{c}\partial_\xi\right){u}(\xi,t)=v(\xi,t) \ ,\\
\label{eq:D_EOM_co}
&&H \Gb\left(c^{-2}\partial_{tt}-2 \beta c^{-1}\partial_{\xi t}-\left(1-\beta ^2\right)\partial_{\xi\xi}\right)u(\xi,t)=\nonumber\\
&&\qquad\qquad\tau_{\rm d}-\sigma{f}\left[v(\xi,t),\phi(\xi,t)\right]\ , \\
\label{eq:D_phi_co}
&&\left(\partial_t\!-\!\beta{c}\partial_\xi\right)\phi(\xi,t)=g\left[v(\xi,t),\phi(\xi,t)\right]\ .
\end{eqnarray}
To derive steady-state solutions, we omit all partial time-derivatives from Eqs.~\eqref{eq:v_co}-\eqref{eq:D_phi_co}. Equation~\eqref{eq:v_co} leads to
\begin{equation}
-\beta{c}\partial_\xi{u}(\xi)=v(\xi)\quad\Rightarrow\quad u(\xi )=-\frac{1}{\beta  c}\int_0^{\xi } v(\Xi) \, d\Xi\ .
\end{equation}
Omitting the time-dependence from Eqs.~\eqref{eq:D_EOM_co}-\eqref{eq:D_phi_co} and substituting the expression for $u$ into Eq.~\eqref{eq:D_EOM_co}, we end up with two ODE's
\begin{align}
\label{eq:D_EOM_ss}
v'(\xi)&=\frac{\beta  c}{\left(1-\beta ^2\right)H\Gb}\left(\tau_{\rm d}-\sigma{f}\left[v(\xi),\phi(\xi)\right]\right)\ , \\
\label{eq:D_phi_ss}
\phi'(\xi)&=-\beta^{-1}c^{-1}g\left[v(\xi),\phi(\xi)\right]\ ,
\end{align}
where $\beta$, the nonlinear eigenvalue in the problem, is still unknown. This dynamical system can be analyzed using standard tools~\cite{SMMeiss2007,SMStrogatz2014}. That being stated, we do note that Eqs.~\eqref{eq:D_EOM_ss}-\eqref{eq:D_phi_ss} are generically stiff in the mathematical sense, i.e.~they feature widely different scales of variation, and hence even their numerical analysis is nontrivial. This is demonstrated explicitly below.

We first look for the fixed-points of Eqs.~\eqref{eq:D_EOM_ss}-\eqref{eq:D_phi_ss}, i.e.~$v'(\xi )\=\phi'(\xi)\=0$, which leads to $\left\{v,\phi\right\}\=\left\{V,\phi_{\rm ss}(V)\right\}$ that satisfy $g\left[V,\phi_{\rm ss}(V)\right]=0$ and $f_{\rm ss}(V)\!\equiv\!f\left[V,\phi_{\rm ss}(V)\right]\=\tau_{\rm d}/\sigma$. Recall that $f_{SS}(V)$ is shown in Fig.~1 in the manuscript. Since we are looking for solutions which approach prescribed values of $V$ as $\xi\!\to\!\pm\infty$, we are interested in the asymptotic behavior of the equations around these fixed-points. A linear stability analysis of Eqs.~\eqref{eq:D_EOM_ss}-\eqref{eq:D_phi_ss} shows that for $f'_{\rm ss}(V)\!>0\!$, i.e.~for $V$ on a velocity-strengthening branch of $f_{\rm ss}(V)$, the fixed-points are in fact saddle points. That is, both eigenvalues are real, one is negative, $\lambda_-\!<\!0$ and one is positive, $\lambda_+\!>\!0$. The associated eigenvectors are denoted by $\vec{q}_\pm$. This implies that any solution converging to a fixed-point as $\xi\!\to\!\pm\infty$ must do so along a particular eigenvector, i.e.~$\vec{q}_-$ for $\xi\!\to\!+\infty$ ($\lambda_-\!<\!0$ is relevant here in order to avoid divergence in this limit) and $\vec{q}_+$ for $\xi\!\to\!-\infty$ ($\lambda_+\!>\!0$ is relevant here in order to avoid divergence in this limit). With these properties in mind, finding steady-state solutions is straightforward. We integrate Eqs.~\eqref{eq:D_EOM_ss}-\eqref{eq:D_phi_ss} starting from both large positive and negative values of $\xi$, using the initial condition $\left\{v,\phi\right\}\=\left\{V,\phi_{\rm ss}(V)\right\}+\varepsilon\vec{q}_\pm$, where $\varepsilon$ is a small parameter. We use an initial guess for the value of $\beta$, which is improved iteratively using a shooting method~\cite{SMPress2007} (similar to that used in~\cite{SMBar-Sinai2012,SMBar-Sinai2013}) until the solutions from both ends meet.

\begin{figure}[ht]
  \centering
  \includegraphics[width=0.90\columnwidth]{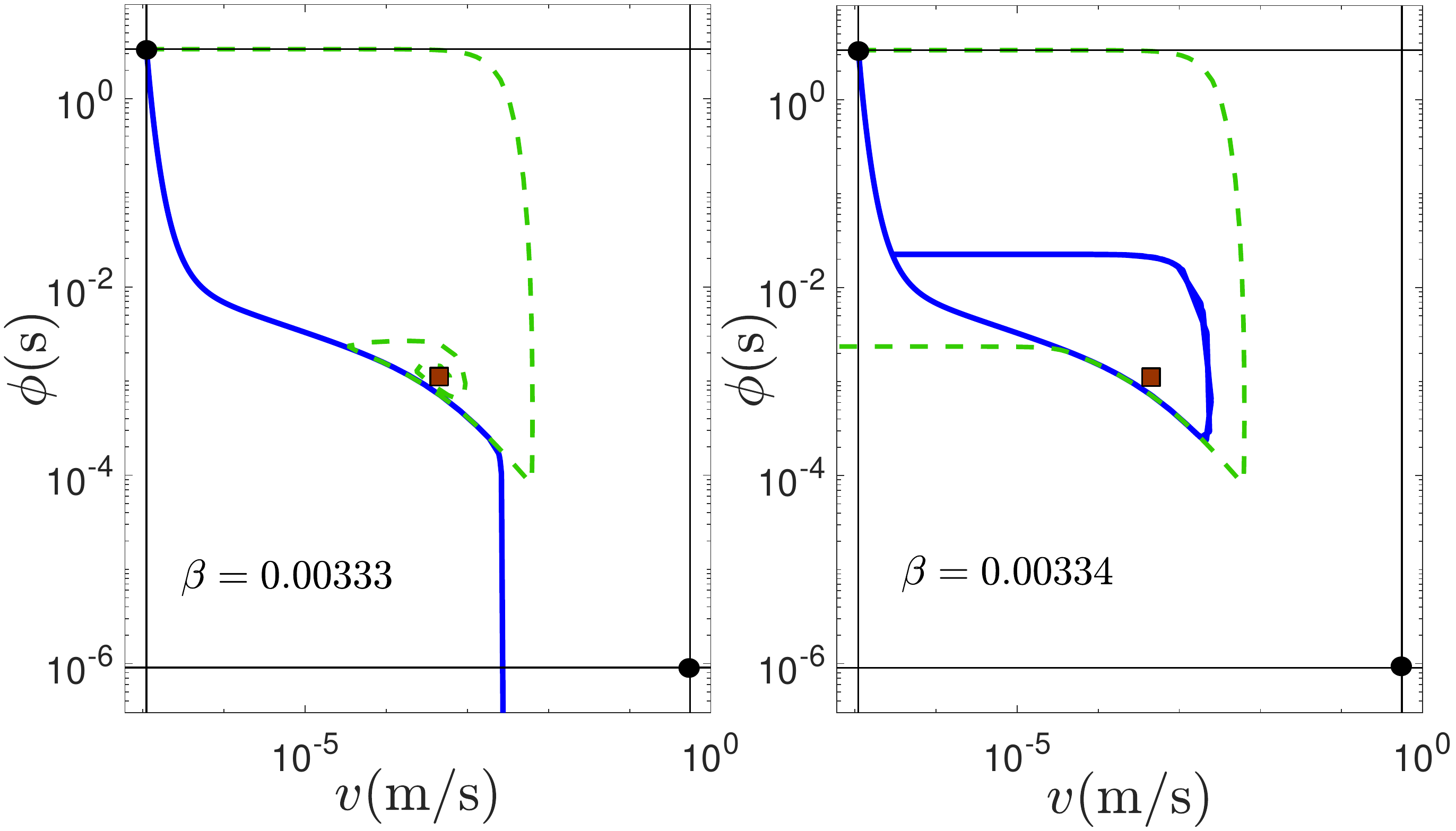}
  \caption{An example of the existence of a steady-state pulse solution (see details in the text). The left panel is for $\beta\!=\!0.00333$ and the right is for $\beta\!=\!0.00334$.}
  \label{fig:phase_pulse}
\end{figure}

An example for the existence of steady-state pulse solutions and their properties, here for $\tau\=1.05\tau_{\rm d}$, is presented in Fig.~\ref{fig:phase_pulse}. Here, as we are looking for a slip pulse solution, we demand that both ends of the solution converge to $\Vst$. The behavior of the system of equations in the $\phi\!-\!v$ plane, for two close values of $\beta$, is considered:
\begin{enumerate}[leftmargin=*]
  \item {\bf Dashed green line on the left panel}: Here we start integrating from the $\Vst$ fixed-point (black circle) along a trajectory that corresponds to a decreasing $\xi$ (starting from a large positive value). The solution first progresses horizontally and then curves down toward the steady-state solution which corresponds to $V_{\rm vw}$ (brown square). For the value of $\beta$ we use here ($\beta\=0.00333$), the $V_{\rm vw}$ fixed-point is a repeller (for an increasing $\xi$) and consequently the solution spirals down toward it (because it is a backwards solution, i.e.~$\xi$ is decreased).
  \item {\bf Solid blue line on the left panel}: Here we start integrating from the $\Vst$ fixed-point (black circle) along a trajectory that corresponds to an increasing $\xi$ (starting from a large negative value). This solution curves down toward $V_{\rm vw}$, but avoids it as it is a repeller.
  \item {\bf Dashed green line on the right panel}: Here the value of $\beta$ is slightly increased ($\beta\=0.00334$). We then start integrating from the $\Vst$ fixed-point (black circle) along a trajectory that corresponds to a decreasing $\xi$ (starting from a large positive value). The solution initially follows the dashed green line of the left panel, but as it approaches $V_{\rm vw}$, it is repelled away from it.
  \item {\bf Solid blue line on the right panel}: The solution initially follows the solid blue line on the left panel, but eventually settles into a finite limit cycle around $V_{\rm vw}$.
\end{enumerate}

\begin{figure}[ht]
  \centering
  \includegraphics[width=0.9\columnwidth]{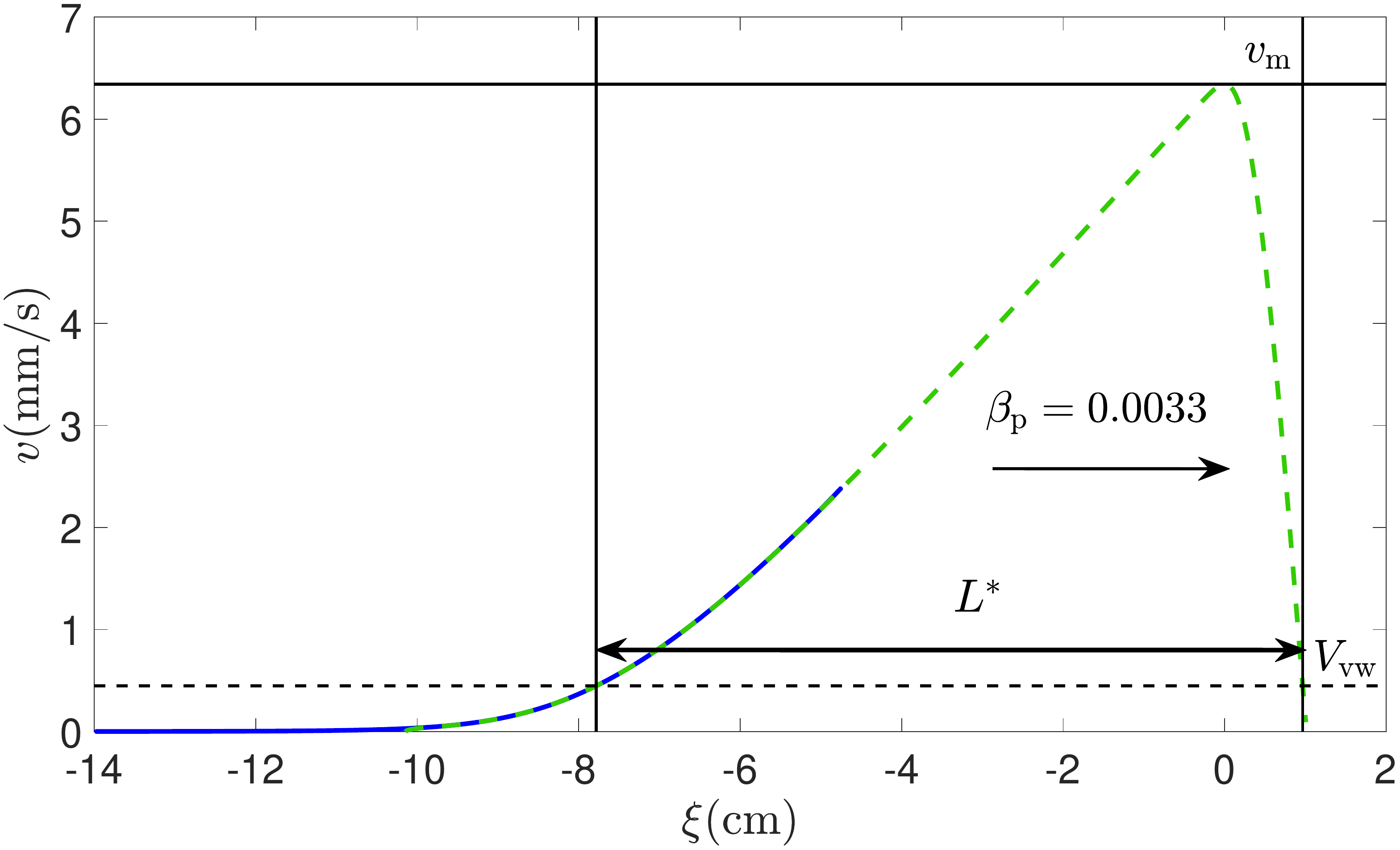}
  \caption{The same as Fig.~2 in the manuscript (here the dimensional quantities are plotted), except that the two segments corresponding to Fig.~\ref{fig:phase_pulse}, are plotted (dashed green and solid blue lines). An almost perfect overlap between the two segments over some spatial range is observed, which justifies plotting a single curve in Fig.~2 in the manuscript.}\label{fig:xi_pulse}
\end{figure}

While the results presented in Fig.~\ref{fig:phase_pulse} do not explicitly demonstrate a steady-state pulse solution, i.e.~a trajectory that starts {\em and} ends at $\Vst$, they do demonstrate without doubt that such a solution exists within an extremely narrow range of $\beta$ values (here between $\beta\!=\!0.00333$ and $\beta\!=\!0.00334$), which is a manifestation of the stiffness of the underlying equations. To see this, note that phase-plane curves change smoothly as a parameter (in our case $\beta$) is varied smoothly. Consequently, there must exist a value of $\beta$ between $\beta\!=\!0.00333$ and $\beta\!=\!0.00334$ for which the qualitatively different behaviors presented in Fig.~\ref{fig:phase_pulse} are exchanged. At this value of $\beta$, which corresponds to a homoclinic bifurcation~\cite{SMMeiss2007,SMStrogatz2014}, a homoclinic solution exists, i.e.~a solution which starts and ends at the same fixed-point, corresponding to a slip pulse. In fact, the slip pulse solution serves as a separatrix~\cite{SMMeiss2007,SMStrogatz2014} between the two qualitatively different behaviors. The pulse propagation velocity can be estimated as the average between $\beta\!=\!0.00333$ and $\beta\!=\!0.00334$ (note though that the exact pulse propagation velocity does not identify with the value of $\beta$ for which $V_{\rm vw}$ changes from a repeller to an attractor). The spatial profile of pulse's slip velocity, as shown in Fig.~2 in the manuscript, can be constructed if the trajectories nearly overlap (to an arbitrary accuracy), as in our case. This is demonstrated in Fig.~\ref{fig:xi_pulse}. From the spatial profile, we can calculate the width of the pulse, $L^*$, defined as the distance between the two points where $v(\xi)\=V_{\rm vw}$. We also calculate $v_{\rm m}$, which is the maximal velocity of the pulse. The procedure described in this example can be repeated for different $\tau_d$'s to obtain the complete spectrum of steady-state slip pulses. A similar procedure is used to derive rupture and healing front solutions.

As described in the manuscript, we next addressed the stability of the steady-state pulse solutions. This is done by solving the partial differential Eqs.~\eqref{eq:v_co}-\eqref{eq:D_phi_co} through the method of lines~\cite{SMSchiesser1991} and using the perturbation procedure described in the manuscript.

\subsection{Steady-state slip pulses as critical nuclei: Perturbations in dynamical calculations}
\label{sec:nuc}

To test the idea that the steady-state slip pulses play the role of non-equilibrium critical nuclei for the onset of rapid slip, we introduced perturbations into the original equations as initial conditions. In 1D, these perturbations are obtained from the steady-state pulse solutions corresponding to $L^*(\tau_{\rm d})$ in the following manner: (i) For $\tau_d\!>\!\tau^*$, perturbations with $L\!>\!L^*$ are obtained by stretching the steady-state pulse solutions corresponding to a given $\tau_{\rm d}$ and perturbations with $L\!<\!L^*$ are obtained by compressing them. (ii) For $\tau_d\!<\!\tau^*$, we used the steady-state pulse solutions corresponding to $L\=L^*$ as initial conditions, but solved the dynamical equation with values in the range $\tau_d\!<\!\tau^*$. We then tracked the system's evolution, by solving Eqs.~\eqref{eq:v}-\eqref{eq:D_phi} using the same method of lines~\cite{SMSchiesser1991} mentioned above, for each point in the $L-\tau_{\rm d}$ plane to determine whether the perturbations decay back to $V_{\rm stick}$ or bring the system to $V_{\rm vs}$, resulting in Fig.~3a in the manuscript.

\begin{figure}[ht]
  \centering
  \includegraphics[width=0.98\columnwidth]{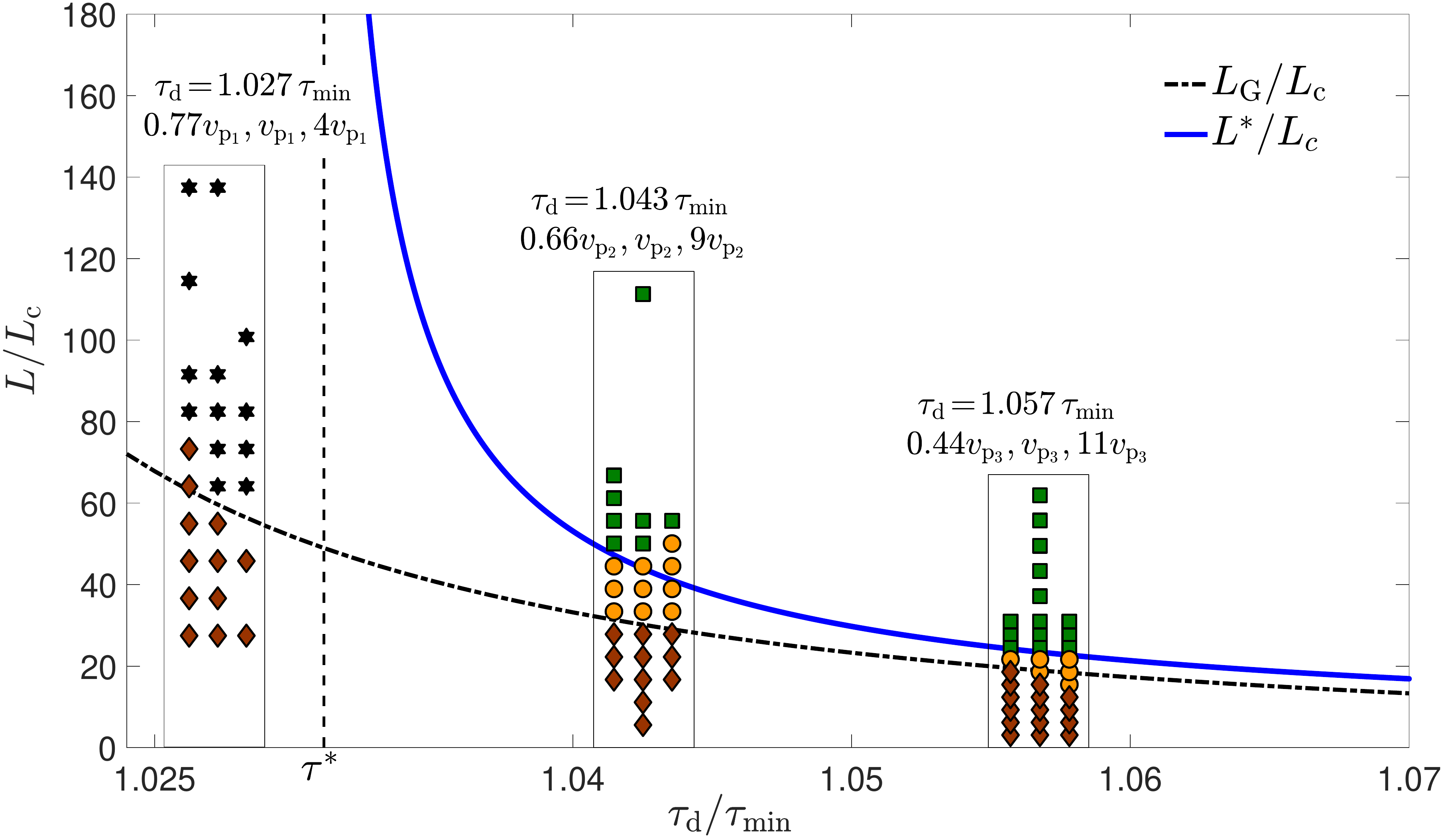}
  \caption{To demonstrate the robustness of our physical picture against variations in the amplitude of the Gaussian perturbations, we present here results for amplitudes both significantly smaller (yet larger than $V_{\rm vw}$) and larger than the amplitude used in Fig.~5 in the manuscript. In particular, for three values of $\tau_{\rm d}$, indicated above each rectangle inside the figure, we used widely different amplitudes (as also indicated above each rectangle, the first one corresponds to the left column, the second one to the middle column and the third one to the right column). The reference amplitude, corresponding to the middle column in each rectangle, is the one used in Fig.~5 in the manuscript (the reference amplitudes are $v_{\rm p_1}\!=\!0.167v_{\rm min}$ and $v_{\rm p_2}\!=\!v_{\rm p_3}\!=\!0.193v_{\rm min}$). The results (using the same symbol and color codes as in Fig.~5) exhibit negligible variations with the amplitude, further demonstrating the robustness of the 2D phase-diagram using Gaussian perturbations.}\label{fig:SSpulse_IC}
\end{figure}

In 2D, we introduced Gaussian perturbations into a steady sliding state at $\Vst$. These perturbations are characterized by a width $L$, corresponding to $10$ Gaussian standard deviations (this choice is explained below), and peak amplitude $v_{\rm p}$. The perturbation in $\phi$ is determined by $v$ through steady-state conditions. The explicit initial conditions take the form
\begin{eqnarray}
v(x,0)&=&\Vst+e^{-\frac{50 x^2}{L^2}} \left(v_{\rm p}-\Vst\right)\ ,\\
u(x,0)&=&0\ ,\qquad\phi(x,0) = \phi_{\rm ss}(v(x,0))\ ,
\end{eqnarray}
and are introduced at the center of the domain. We used $v_{\rm p}\=0.167v_{\rm min}$ (except for the smallest $\tau_{\rm d}$ value for which we used $0.193v_{\rm min}$). The 2D equations, to be described in more detail in Sect.~\ref{sec:2D} below, were solved, leading to Fig.~5 in the manuscript. The robustness of the 2D phase-diagram against variations in the amplitude of the Gaussian perturbations, as long as it is larger than $V_{\rm vw}$, is demonstrated in Fig.~\ref{fig:SSpulse_IC}.

\subsection{Steady-state slip pulses for friction curves that feature no minimum}
\label{sec:VW}

\begin{figure}[ht]
  \centering
  \includegraphics[width=0.9\columnwidth]{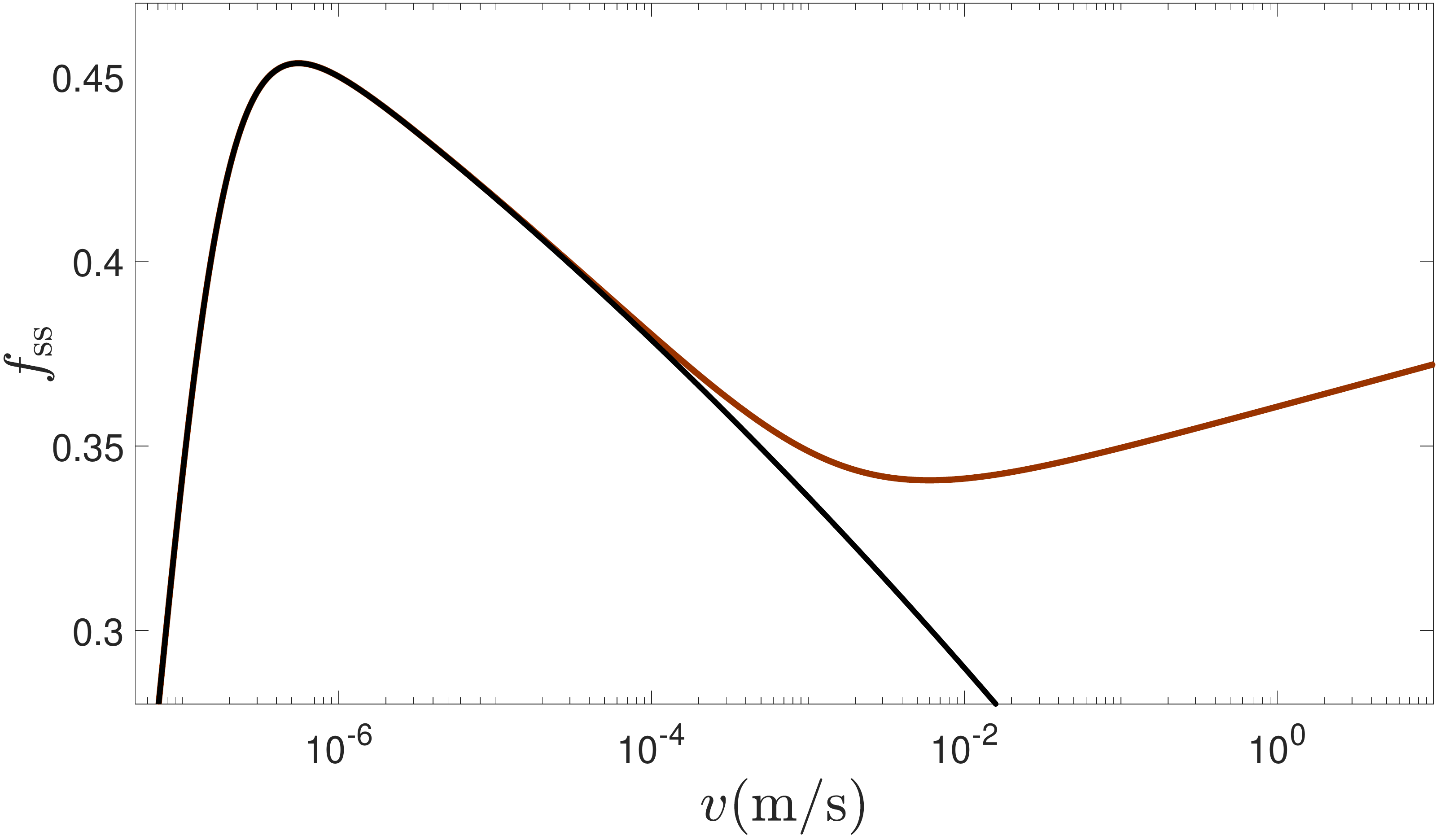}
  \caption{The steady-state friction coefficient, $f_{\rm ss}$, corresponding to steady-state conditions in Eq.~\eqref{eq:f} (brown line, which is identical to the solid brown line in Fig.~1 in the manuscript). The same, except that the ``1'' in the $\phi$ logarithmic term in Eq.~\eqref{eq:f} is omitted (black line), resulting in the elimination of the minimum.}
  \label{fig:no-minimum}
\end{figure}

\begin{figure}[ht]
  \centering
  \includegraphics[width=0.94\columnwidth]{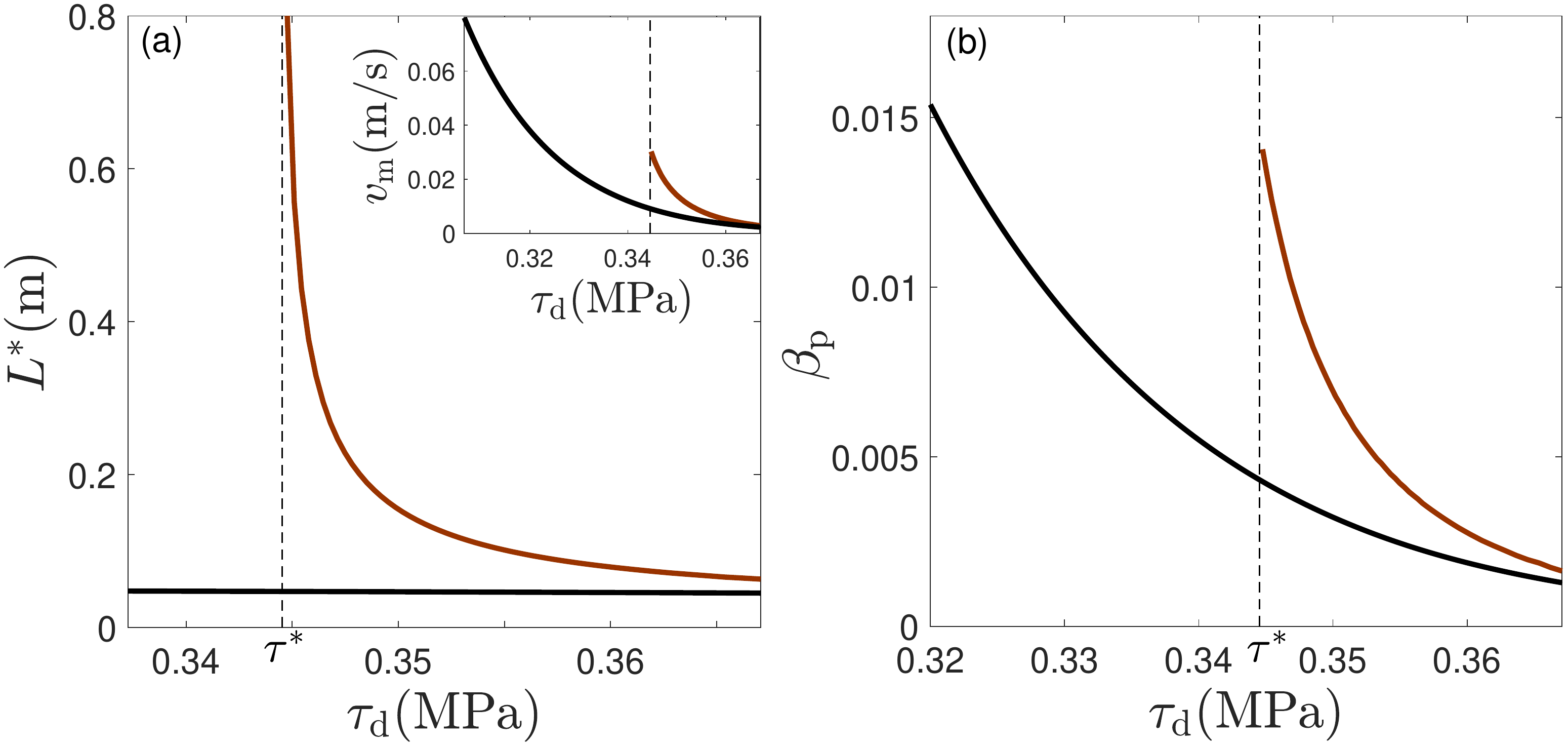}
  \caption{$L^*(\tau_{\rm d})$ (panel a), $v_{\rm m}(\tau_{\rm d})$ (inset panel a) and $\beta_{\rm p}(\tau_{\rm d})$ (panel b) for both the friction curve with a minimum (brown lines here and in Fig.~\ref{fig:no-minimum}) and without (black lines here and in Fig.~\ref{fig:no-minimum}). The brown lines here are identical to the blue lines in Fig.~3 in the manuscript, just without any normalization. The dashed vertical line represents $\tau^*$.}\label{fig:VW_prop}
\end{figure}

As argued in the manuscript, while it is appealing to think of steady-state slip pulses as emerging from the interaction of steady-state rupture and healing fronts, which vanishes at a finite stress $\tau^*$, this is not a necessary condition. To show this, we consider a steady-state friction curve that features no minimum, for which steady-state rupture and healing fronts --- and hence $\tau^*$ --- do not exist. In particular, we consider $f$ given in Eq.~\eqref{eq:f} above, except that we drop the ``1'' in the $\phi$ logarithmic term. This change (the $\dot\phi$ equation remains unchanged) eliminates the minimum~\cite{SMBar-Sinai2012,SMBar-Sinai2013}, as shown in Fig.~\ref{fig:no-minimum}. We repeated the calculations described above and derived steady-state slip pulse solutions in this case. The results for $L^*(\tau_d)$, $v_{\rm m}(\tau_d)$ and $\beta_{\rm p}$ are shown in Fig.~\ref{fig:VW_prop}, together with the results reported in Fig.~3 in the manuscript. We observe that steady-state slip pulses exist in the absence of a minimum in the steady-state friction curve, where the main difference relative to the case in which a minimum exists is the absence of a finite $\tau^*$.

\newpage

We note in passing that in 2D (and 3D) in infinite systems the radiation of elastic waves from the interface to infinity effectively alters the friction law such that an effective minimum in the steady-state friction relation emerges, even when the bare relation features no minimum. In particular, this radiation process appears as a contribution that is proportional to the slip velocity in the interfacial relation, cf.~Eq.~\eqref{spectral} and the text below it. Consequently, we expect such 2D (and 3D) calculations --- even in the absence of pure interfacial velocity-strengthening contribution to the bare friction law --- to yield qualitatively similar results to ours. Indeed, in~\cite{SMPerrin1995,SMZheng1998} such calculations have been performed in the context of searching for self-healing slip pulses; the results indicate the existence of sustained pulses under certain applied stress conditions, which appears to be at least qualitatively consistent with our findings in this context (see Fig.~5 in the manuscript).

\subsection{Nucleation in 2D infinite systems under anti-plane shear conditions}
\label{sec:2D}

As explained in the manuscript, we performed 2D mode-III elastodynamic calculations using the spectral boundary integral formulation~\cite{SMGeubelle1995,SMMorrissey1997,SMBreitenfeld1998}. The latter relates the traction stresses acting along the interface between two semi-infinite linearly elastic
half-spaces and the resulting displacements. For the mode-III elastodynamic
problem studied in the manuscript, the interface is initially uniformly
pre-stressed by $\tau_{\rm d}$ and set to slide at an extremely small steady velocity
$V_{\rm stick}$, such that the shear tractions at the interface take the
form
\begin{equation}
\tau(x,t) = \tau_{\rm d} -\frac{\mu}{2c_s}\Big(v(x,t)-V_{\rm stick}\Big) +
s(x,t) \ .
\label{spectral}
\end{equation}
The second right-hand side term represents the instantaneous response to a
change in the sliding velocity, the so-called radiation damping term. This term can be understood as the damping of interfacial energy
due to elastic waves radiated into the infinite domain. The third term $s(x,t)$ accounts for the
history of interfacial displacements. Both $s(x,t)$ and $u_z(x,t)$ are related in
the spectral domain via a convolution integral
\begin{equation}
S(k,t) = -\mu|k|\int_0^t H\Big(|k|c_s(t-t')\Big)U_z(k,t')|k|c_s\:dt' \ ,
\label{convolution}
\end{equation}
where $S(k,t)$ and $U_z(k,t)$ are the spatial Fourier transforms of $s(x,t)$ and $u_z(x,t)$, respectively.

In Eq.~\eqref{convolution}, the convolution kernel $H(\gamma)$ (not to be confused with the finite system height of previous sections) is expressed from the
Bessel function of the first kind $J_1(\gamma)$ as
\begin{equation}
H(\gamma)=\gamma^{-1}J_1(\gamma) \ .
\end{equation}
Due to the spectral nature of the formulation, the simulated domain is taken to be periodic,
with periodicity $X$. The latter is chosen to be large enough to prevent
any effect of the periodicity on the results reported in the phase-diagram of
Fig.~5. The sliding velocity is computed by combining Eq.~\eqref{spectral} and the
friction law $\tau=\sigma\,\mathrm{sgn}(v)f(|v|,\phi)$. $u_z$ and $\phi$ are
then integrated in time using an explicit time-stepping scheme
\begin{eqnarray}
\label{eq:tstep1}
&\hspace{-1cm}u_z(x,t+\Delta t) = u_z(x,t) + 0.5 v(x,t)\Delta t,\\
\label{eq:tstep2}
&\hspace{-1cm}\phi(x,t+\Delta t) = \phi(x,t) + g\Big(|v(x,t)|,\phi(x,t)\Big)\Delta t \ .
\end{eqnarray}
Note that the factor $0.5$ on the right-hand-side of Eq.~\eqref{eq:tstep1} ensures that $v(x,t)$ is indeed the slip velocity.
In order to guarantee the stability and the convergence of the numerical
scheme, $\Delta t$ is defined as the time needed for a shear wave to
travel a fraction $\delta\=0.2$ of one grid spacing, i.e.~$\Delta t\=\delta\Delta x/c_s$.

In Fig.~5 in the manuscript, the Griffith-like length (see manuscript for details)
\begin{equation}
\label{eq:L_G}
L_{\rm G}=4\mu\pi^{-1} G_{\rm c}\left(\tau_{\rm d}-\tau_{\rm res}\right)^{-2}\ ,
\end{equation}
is plotted (dashed-dotted line). This is made possible once the residual shear stress behind the crack tip during the initial stages of nucleation, $\tau_{\rm res}$, and the effective fracture energy, $G_{\rm c}$, are calculated. The former, which appears to be nearly constant for a broad range of conditions, is directly extracted from the numerical calculations and takes the value $\tau_{\rm res}\=0.34$MPa. To calculate $G_{\rm c}$, we briefly describe a procedure to be discussed in details in~\cite{SMBarras2018}.
\begin{figure}[ht]
  \centering
  \includegraphics[width=\columnwidth]{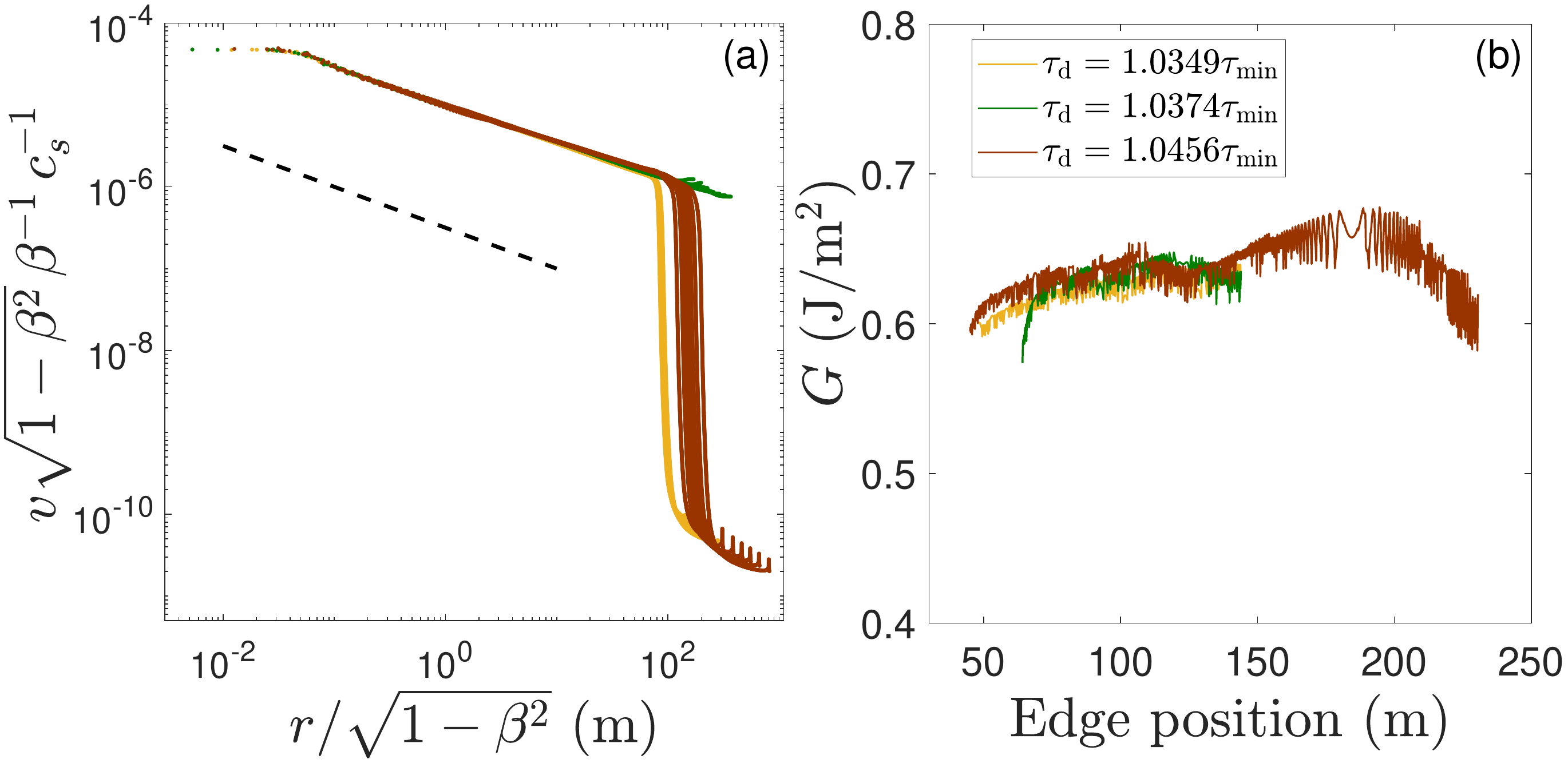}
  \caption{(a) A normalized $v(r,t)$ vs.~a normalized $r$ for a rupture front (green) and two slip pulses (orange and brown) for several times $t$. The dashed black has a $-1/2$ slope, directly demonstrating the validity of  the intermediate asymptotic relation in Eq.~\eqref{eq:v_K}. (b) The corresponding energy release $G$, calculated through Eq.~\eqref{eq:G_K} once $K_{III}$ is extracted from (a), vs.~the edge position.}
  \label{fig:G}
\end{figure}

First, note that energy balance near the edge of a front/pulse reads $G_{\rm c}\=G$, where $G$ is the linear elastic energy release rate~\cite{SMfreund1998dynamic}. $G$ is associated with a near-edge crack-like square root singularity, whose intensity is quantified by the stress intensity factor, $K_{III}$~\cite{SMfreund1998dynamic}. In particular, $G$ is related to $K_{III}$ through~\cite{SMfreund1998dynamic}
\begin{equation}
\label{eq:G_K}
G=\frac{K_{III}^2}{2\mu\sqrt{1-\beta^2}}\ ,
\end{equation}
where $\beta$, as in the manuscript, is the dimensionless front/pulse propagation velocity (here in units of the shear wave speed $c_s\=\sqrt{\mu/\rho}$, where $\rho$ is the mass density). We extract $K_{III}$ from the intermediate asymptotic velocity behind the edge, given by~\cite{SMfreund1998dynamic}
\begin{equation}
\label{eq:v_K}
v(r,t)\simeq\frac{2 \beta c_s K_{III}(t)}{\mu\sqrt{2 \pi\left(1-\beta ^2\right) r}}\ ,
\end{equation}
where $r$ is the distance from the edge. The left panel of Fig.~\ref{fig:G} shows three examples of $v(r,t)$, one corresponding to a rupture front (green curves) and two to slip pulses (orange and brown), each features a series of snapshots in time.

In all examples, there exists a spatial range where $v\!\propto\!1/\sqrt{r}$ (black dashed line, a guide to the eye), as expected. By extracting $K_{III}(t)$ from these curves, we can calculate $G(t)$ through Eq.~\eqref{eq:G_K}. The result, where $G$ is plotted as a function of the front/pulse edge position (not time $t$), is shown on the right panel of Fig.~\ref{fig:G}. We observe that $G$ varies only slightly during propagation, and hence we approximate it by a constant $G_{\rm c}\=0.65$J/m$^2$, used to calculate $L_{\rm G}$ through Eq.~\eqref{eq:L_G}. We note that while an edge singularity does not exist in 1D, global energy balance considerations can still be used to define a Griffith-like length, which in this case scales as $\sqrt{\Gb\,H\,G_c}\,(\tau_{\rm d}-\tau_{\rm res})^{-1}$~\cite{SMPuzrin2005}. We extensively searched for signatures of the 1D Griffith-like length in our 1D analysis, but found none.

The results of our 2D mode-III calculations are summarized in the phase-diagram in Fig.~5 in the manuscript. The Gaussian perturbations procedure is described above in Sect.~\ref{sec:nuc}. Note that the width of perturbations $L$ is defined to correspond to $10$ standard deviations so that $L_{\rm G}$ indeed separates decaying from propagating perturbations (see manuscript for details) for a {\em single} value of $\tau_{\rm d}$. Finally, the phase-diagram features $4$ different dynamical behaviors (``phases''), denoted by different symbols and colors in Fig.~5 in the manuscript. We verified that the phase-diagram is independent of the amplitude of perturbations, as long as it is larger than $V_{\rm vw}$, see Fig.~\ref{fig:SSpulse_IC}. We attach to this Supplemental Material $4$ movies, each corresponds to a different dynamical behavior (``phase''), according to the following list:
\begin{itemize}[leftmargin=*]
\item \textbf{Movie\_S1:} Decay without propagation arising for $L\!<\!L_{\rm G}$ (brown diamonds in Fig.~5).
\item \textbf{Movie\_S2:} Nucleation arising for $L\!>\!L^*$ (green squares in Fig.~5).
\item \textbf{Movie\_S3:} Decay with propagation arising for $L\!>\!L_{\rm G}$ and $\tau_{\rm d}\!<\!\tau^*$ (black hexagrams in Fig.~5).
\item \textbf{Movie\_S4:} Sustained pulses arising for $L_{\rm G}\!<\!L\!<\!L^*$ and $\tau_{\rm d}\!>\!\tau^*$ (orange circles in Fig.~5).
\end{itemize}

Note that in the movies $V_{\rm vw}$ and $V_{\rm vs}$ correspond to the steady-state solutions of Eq.~\eqref{spectral}, i.e.~$\sigma f(v,\phi\=D/v)\=\tau_{\rm d} -\frac{\mu}{2c_s}(v-V_{\rm stick})$, which do not identify with those defined in Fig.~1 in the manuscript.

Finally, we note that propagation velocities, both of rupture fronts and of pulses, are generally larger in 2D compared to 1D. To see this, note that Eq.~\eqref{eq:phi} is dimension-independent and takes the form $\beta\!\sim\!\frac{L v}{D c_s}$, where $L$ is the characteristic size and $v$ is the characteristic slip rate. $L$ is known to increase with the system size $H$ before it saturates to an $H$-independent value (in the small $H$ regime it scales with $\sqrt{H}$)  and hence is larger in 2D than in 1D. In addition, the slip rate $v$ is expected to be larger in 2D in the presence of a crack-like tip singularity (which is absent in 1D). Finally, the presence of velocity-strengthening friction also affects the propagation speed. In particular, in the absence of velocity-strengthening friction, as discussed in Section~\ref{sec:VW} above, both the maximal slip rate (inset of Fig.~\ref{fig:VW_prop}a) and the propagation velocity (Fig.~\ref{fig:VW_prop}b, note as $\tau^*$ does not exist for pure velocity-weakening friction, decreasing the driving stress $\tau_{\rm d}$ will lead to an increase in the propagation velocity) increase.

\subsection{Parameters}

The parameters used for all the calculations described in the manuscript and here are given in Table~\ref{tab:values}.
\begin{table}[ht]
  \centering
  \begin{tabular}{|c|c|c|}
  \hline
  Parameter & Value & Units\\
  \hline
  $\Gb,\mu$ & $9\!\times\!10^9$ & Pa\\ \hline
  $H$ & $2\!\times\!10^{-4}$ &m \\ \hline
  $\sigma$ & $10^6$ & Pa\\ \hline
  $c,\,c_s$ & $2739$ & m/s\\ \hline
  $D$ & $5\!\times\!10^{-7}$ &m \\ \hline
  $b$ & $0.075$ & -\\ \hline
  $v_*$ & $10^{-7}$ & m/s\\ \hline
  $f_0$ & $0.28$ & -\\ \hline
  $\phi_*$ & $3.3\!\times\!10^{-4}$ & s\\ \hline
  $\alpha$ & $0.005$ & -\\
   \hline
\end{tabular}
  \caption{Values for all parameters used (in MKS units).}\label{tab:values}
\end{table}
Note that the values of the listed parameters are characteristic of some laboratory experiments (see~\cite{SMBar-Sinai2014} for details). However, the generic properties of the derived results are independent of the exact numbers, and are relevant to a broad range of materials and physical situations. For example, $v_*$ that controls the velocity scale below which the system is in the stick phase, can be taken to be significantly smaller, and larger confining pressures $\sigma$ can be considered.


%

\end{document}